# Dynamic Cooperative Strategies in Search Engine Advertising Market: With and Without Retail Competition

Huiran Li[1], Qiucheng Li[1], Baozhu Feng[2,*]

[1]*School of Business Administration and Customs Affair, Shanghai Customs College, Shanghai 201204, China*

[2]*School of Business Administration, Anhui University of Finance and Economics, BengBu 233030, Anhui, China*

**Abstract:** In search engine advertising (SEA) market, where competition among retailers is intense and multifaceted, channel coordination between retailers and manufacturers emerges as a critical factor, which significantly influences the effectiveness of advertising strategies. This research attempts to provide managerial guidelines for cooperative advertising in the SEA context by modeling two cooperative advertising decision scenarios. Scenario I defines a simple cooperative channel consisting of one manufacturer and one retailer. In Scenario II, we consider a more general setting where there is an independent retailer who competes with the Manufacturer-Retailer alliance in Scenario I. We propose a novel cooperative advertising optimization model, wherein a manufacturer can advertise product directly through SEA campaigns and indirectly by subsidizing its retailer. To highlight the distinctive features of SEA, our model incorporates dynamic quality scores and focuses on a finite time horizon. In each scenario, we provide a feasible equilibrium solution of optimal policies for all members. Subsequently, we conduct numerical experiments to perform sensitivity analysis for both the quality score and gross margin. Additionally, we explore the impact of the initial market share of the competing retailer in Scenario II. Finally, we investigate how retail competition affects the cooperative alliance's optimal strategy and channel performance. Our identified properties derived from the equilibrium and numerical analyses offer crucial insights for participants engaged in cooperative advertising within the SEA market.

# 1. Introduction

Over the past two decades, Search Engine Advertising (SEA) has become a leading force in online advertising (Yang and Li, 2023). According to the Interactive Advertising Bureau (2024), SEA dominates online advertising revenue in the United States, holding the largest market share at 39.5 percent, with revenues totaling $88.8 billion. According to Organic SEO (2024), retailing plays an important role in SEA, exemplified by Amazon's leading position in Google AdWords spending, with an estimated monthly investment of $22 million. This retail dominance is further demonstrated by two direct retail domains (wayfair.com, ebay.com) and five retail-supporting platforms ranking among the top 10 Google AdWords spenders. In SEA market, consumers search for specific keywords on search engines (e.g., Google), while advertisers, including both manufacturers (e.g., brand owners) and retailers, bid on these keywords to display their advertising alongside organic search results (Bhattacharya et al., 2022). As illustrated in Figure 1, which shows the search results for the query "iphone" on Google, manufacturers can advertise their products (e.g., iPhone) through both their own e-commerce sites (i.e., Apple) and retail partners (e.g., Amazon). This scenario reflects the strategic coordination between manufacturers and retailers, where manufacturers often subsidize retailers' advertising expenditures through cooperative advertising in SEA.

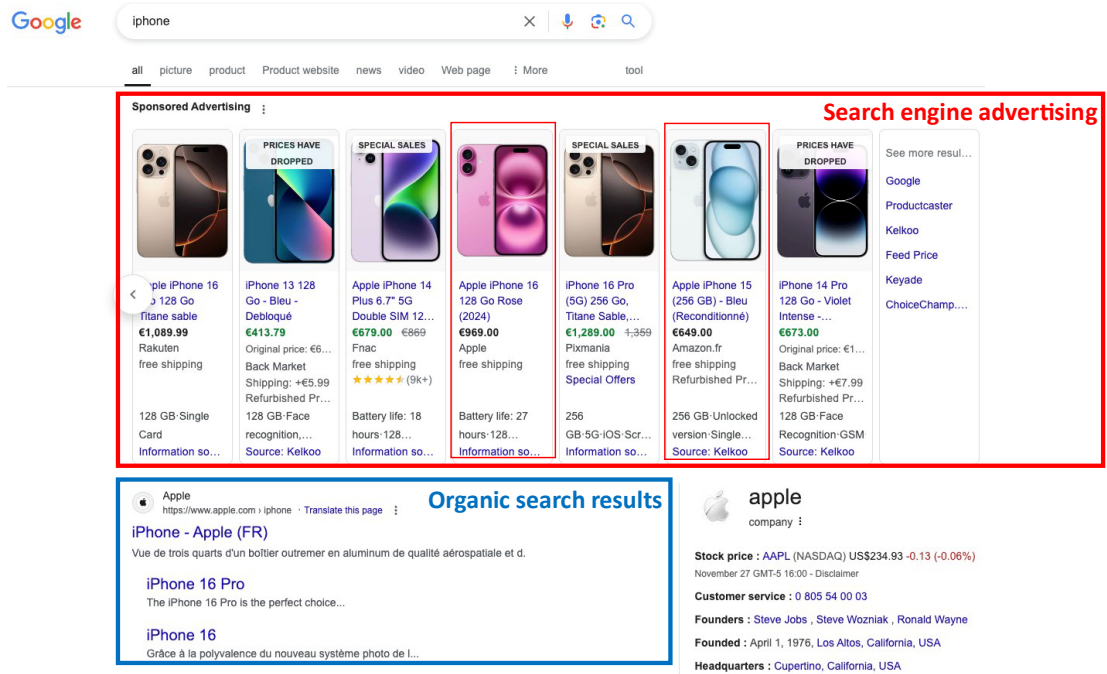

Figure 1. An example of Google search results for "iphone".

Before its integration into SEA, cooperative advertising has long been established in traditional marketing channels. For example, Dove, a personal care brand, provides advertising funds to Walmart, a major retail chain, for creating in-store displays and joint advertisements. This cooperation enables Dove to reach a broad audience through Walmart's extensive customer base, while Walmart benefits from its association with a trusted brand (EMB Global, 2024). Similarly, Intel shares 40% of advertising costs with its supply chain partners who incorporate Intel's image or logo in their advertisements (He et al., 2019). As marketing channels transition into the digital era, cooperative advertising has evolved to embrace online platforms. Rebecca Lieb, founding partner at Kaleido Insights, describes this transformation as a "phenomenal opportunity". This trend is reflected in the substantial market volume, with cooperative advertising spending reaching $50 billion annually in the United States and $520 billion globally, where the digital segment is projected to rapidly expand to $5-10 billion per year (MIT Sloan Management Review, 2019).

Among these digital channels, search engine has emerged as an exceptionally effective platform for cooperative advertising, offering advantages over traditional advertising through its precise targeting capabilities and enhanced return on investment, facilitated by query-keyword match technology and generalized second-price position auctions (Li and Yang, 2022; Dayanik and Sezer, 2023). This shift to SEA has become increasingly critical for manufacturers seeking visibility in competitive online spaces and retailers managing the high costs of premium advertising placements (ChannelSight, 2024). The implementation of cooperative advertising in SEA yields multiple strategic benefits, including increased sales volume, market expansion, and enhanced consumer feedback mechanisms (Gou et al., 2020; Wu et al., 2023; Martín-Herrán and Sigué, 2023). For example, New Balance's strategic shift from traditional cooperative advertising to SEA campaigns with retailers yielded positive outcomes, with 19% of participating retailers reporting increased website traffic (MIT Sloan Management Review, 2019). These demonstrated benefits underscore the importance for both retailers and manufacturers to develop and optimize effective cooperative advertising strategies in the SEA market.

Cooperative advertising has attracted substantial academic attention, with researchers examining it widely across various dimensions. Studies have investigated different supply chain structures (e.g., multiple manufacturers, multiple retailers) (Liu et al., 2014; Karray et al., 2017; Chutani and Sethi, 2018; Zhang et al., 2020), analyzed various effects (e.g., advertising threshold effect, brand halo effect, and delayed effect) (Zhang et al., 2013; Machowska, 2019; Yu et al., 2021), explored the impact of cooperative advertising on operational decisions (e.g., pricing, inventory management) (Krishnamoorthy et al., 2010; Zhou et al., 2018; De Giovanni et al., 2019), and examined cooperative advertising applications in various contexts (e.g., mobile platforms, online to offline supply chains) (Wang et al., 2016; Li et al., 2019). However, despite these extensive studies, there is a notable gap in the literature regarding cooperative advertising strategies within the SEA context. To the best of our knowledge, only Cao and Ke (2019) have examined cooperative advertising in SEA, focusing specifically on bidding strategies in a static coordination game involving one manufacturer and two retailers in SEA. The core element of cooperative advertising, the manufacturer sharing advertising expenditure to incentivize increased retailer advertising investment, and particularly its key component of optimal advertising effort decisions that directly impact advertising expenditure (Zhang et al., 2017; Wu et al., 2023), remains unexplored in the SEA literature.

This research gap is probably due to the complexity of implementing cooperative advertising in SEA, as evidenced by real-world business data and industry insights. According to MIT Sloan Management Review (2019), 45% of retailers do not use all their available cooperative funds. Industry experts have consistently highlighted the difficulties in transitioning traditional cooperative advertising practices to the digital realm. For example, Ben Carcio, CEO and cofounder of Promoboxx, a marketing platform provider for local retailers, describes cooperative marketing as "a broken process with time-consuming fulfillment methods" that wasn't designed for today's digital-first consumer environment. This perspective is reinforced by Gordon Borrell, CEO of advertising research company Borrell Associates, who states that advertisers are overwhelmed by available options and lack sophisticated digital marketing expertise.

Specifically, the unique nature of SEA presents the following challenges. First, its market is inherently dynamic, characterized by rapid changes in consumer behavior, evolving search engine algorithms, and the continuous emergence of new platforms and technologies (Zhang et al., 2014; Dayanik and Sezer, 2023). However, most research on cooperative advertising strategies employs static models (Zhou et al., 2018; Yan and He, 2020), which often fail to capture the complexity and volatility inherent in the real-world SEA market. Although recent studies have increasingly explored dynamic cooperative advertising strategies (Bai et al., 2018; Ezimadu, 2019), these studies typically presume an infinite planning horizon for the derivation of time-independent stationary solutions, primarily to simplify the technical analysis. This assumption fails to align with practical scenarios in SEA marketing channels, where cooperative advertising agreements are frequently bound by finite-term contracts (Zhang and Feng, 2011; Yang et al., 2015a; Yang et al., 2022). The planning horizon, denoted by $T$, may reflect specific seasonal effects (e.g., in the fashion industry or for perishable products), or it may correspond to a product's obsolescence date or daily budget adjustments by advertisers (Jørgensen and Zaccour, 2014; Pnevmatikos et al., 2018). Therefore, a finite planning horizon framework is essential for accurately analyzing optimal cooperative advertising strategies in SEA. It not only provides a realistic view of market dynamics, but also facilitates informed strategic decision-making in an ever-changing environment.

Another significant challenge in SEA lies in its varying levels of competition (Animesh et al., 2011; Yang et al., 2014; Yang et al., 2015b). As illustrated in Figure 2, search result pages for long-tail keywords (e.g., "homemade coffee filters") show no advertising competition and consist of organic results, whereas popular keywords (e.g., "coffee filters") attract intense competition, where multiple advertisers compete for limited advertising display slots to capture customer attention (Ayanso and Karimi, 2015; Yuan et al., 2017; Google Ads, 2024a). This competition is further influenced by quality score, a unique parameter in SEA that serves as a diagnostic tool to evaluate advertisement quality compared to other advertisers (Google Ads, 2024b). The quality score directly impacts both advertising costs and ad placement, making it a critical factor in cooperative advertising decisions. However, few studies have thoroughly explored either the full range of competition scenarios or the role of quality score in SEA cooperative advertising. Our research aims to fill these gaps by analyzing optimal advertising effort related decisions for cooperative strategies in SEA, with consideration of both competition intensity and quality score impacts. To the best of our knowledge, this is the first research in this direction.

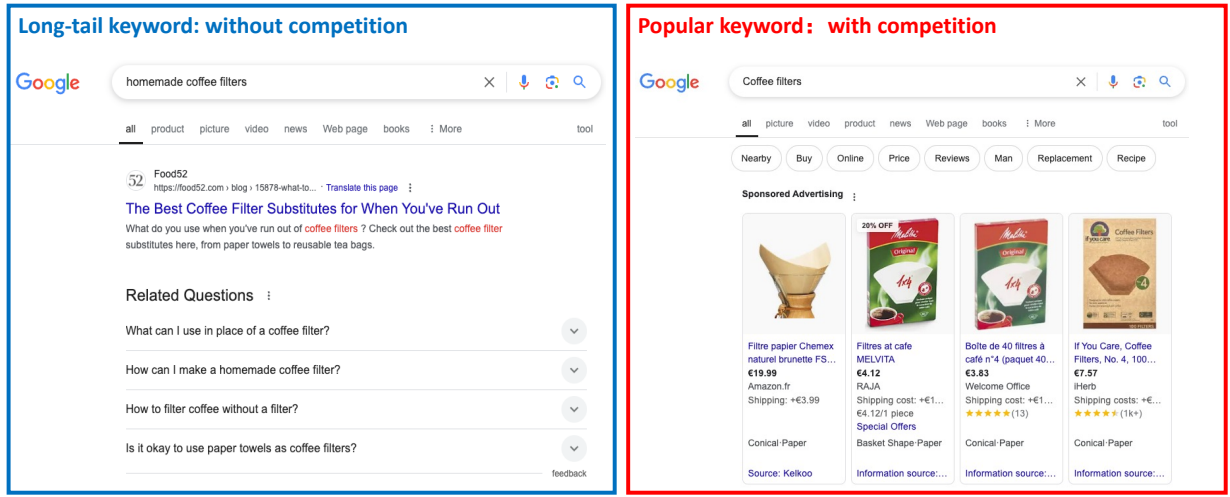

Figure 2. Search engine advertising with and without competition.

In this work, we propose dynamic cooperative advertising optimization models within a finite time horizon, addressing two distinct scenarios: with and without competition. Specifically, Scenario I consists of one manufacturer and one retailer, where the manufacturer distributes products through the retailer. Scenario II adds a competitive independent retailer who competes against the Manufacturer-Retailer alliance established in Scenario I. In both scenarios, the manufacturer can develop its own SEA campaigns and share a percentage of the retailer's advertising expenditure. We propose a feedback equilibrium analysis to determine the optimal advertising efforts for the manufacturer and retailers, and develop a computational solution to derive concrete optimal subsidy rates for the manufacturer in the cooperative advertising. Then, a series of numerical experiments are conducted to evaluate how critical factors in SEA (e.g., quality score, gross margin, and the initial market share) influence the optimal advertising efforts, the corresponding trajectory of market share, and the optimal subsidy rates. Finally, we compared the cooperative alliance's optimal strategy and channel performance between the two scenarios to explore the impact of retail competition.

Experimental results show that: (1) Our approach provides an effective dynamic analytical framework that helps advertisers obtain optimal cooperative strategies within a finite planning horizon, across different competition scenarios in SEA. (2) Quality scores and gross margins play important roles in the cooperative strategies of SEA. Specifically, with higher quality scores, manufacturers tend to achieve larger market shares and higher profits; with higher gross margins, manufacturers are more inclined to increase subsidy rate to the retailer. This increase in subsidy rate motivates the retailer to intensify its advertising efforts, which, in turn, elevates the overall profit of the alliance. (3) Retail competition significantly influences the optimal decisions (i.e., advertising effort and subsidy rate) and channel performance of the cooperative alliance. Improving quality scores emerges as a crucial strategy to enhance its strength in competitive SEA market, thereby gaining more market share and profits.

The remainder of this paper is organized as follows. Section 2 provides a brief literature review. In Section 3, we introduce two cooperative advertising scenarios and construct advertising response functions and profit functions. Section 4 and Section 5 analyze equilibrium solutions and conduct numerical experiments for the two scenarios. Section 6 studies the impact of retail competition. Finally, we conclude this research in Section 7.

## 2. Literature Review

This study primarily focuses on two research streams, i.e., cooperative advertising strategies and advertising competition.

## 2.1 Cooperative advertising strategies

Cooperative advertising strategies have been widely recognized as crucial mechanisms for optimizing supply chain efficiency (Martín-Herrán and Sigué, 2023; Wu et al., 2023). The literature in this domain has evolved through two distinct methodological approaches: static and dynamic modeling. Static models have established the foundational frameworks of cooperative advertising strategies (Ahmadi-Javid and Hoseinpour, 2012; He et al., 2013). For example, Chaab and Rasti-Barzoki (2016) explored supply chain coordination through cooperative advertising and pricing by examining four game structures, including the Nash, Stackelberg retailer, Stackelberg manufacturer, and cooperation games, and extending a consumer demand function for the cooperative advertising model. Jena et al. (2017) developed mathematical models for different closed-loop supply chain configurations, demonstrating that cooperative advertising models yield higher total supply chain profits compared to non-cooperative scenarios. Sarkar et al. (2020) presented a cooperative advertising model considering fuzzy cost variables among suppliers, manufacturers, and retailers, and found that cooperative advertising strategy significantly enhanced overall supply chain revenue. Hong et al. (2023) examined the optimal advertising structure and disclosure strategy for manufacturers in a context of product quality information asymmetry. They discovered that when cooperative advertising is more effective than manufacturer advertising and product quality is high, all channel members can benefit from cooperative advertising.

Recognizing the dynamics of advertising market, researchers have increasingly adopted dynamic models to better address time-dependent effects in supply chain decisions (Zhang et al., 2013; Bai et al., 2018; Han et al., 2023). In dynamic cooperative advertising, the combination of Stackelberg differential game theory with Sethi's stochastic sales-advertising model is widely used.

For example, He et al. (2009) modeled cooperative advertising as a stochastic Stackelberg differential game whose dynamics follows Sethi's stochastic sales-advertising model, incorporating uncertainty in awareness share and deriving optimal feedback policies for pricing and advertising by both the manufacturer and the retailer. To model the effect of subsidies on individual and channel payoffs, Ezimadu (2019) employed Sethi's sales-advertising model to investigate subsidy transfer in cooperative advertising utilizing a differential game framework, where the manufacturer acted as the Stackelberg leader, the distributor and the retailer as the followers. In the exploration of optimal resource allocation within a bilateral monopoly, Karray et al. (2022) utilized a game-theoretic model over a two-period planning horizon. They revealed that manufacturers should avoid from offering exclusively cooperative advertising programs to retailers, based on comparison of equilibrium strategies and profits across various scenarios. Martín-Herrán and Sigué (2023) established a two-period model to examine how members of a bilateral monopoly channel should make pricing and advertising decisions, considering the first-period price as the reference price for the second period, revealing the impact of reference price effects on cooperative advertising decisions in marketing channels.

**2.2 Advertising competition**

The literature on advertising competition reveals its crucial role in shaping optimal advertising strategies, particularly in cooperative advertising contexts (Chutani and Sethi, 2018; Machowska, 2019). Liu et al. (2014) explored the effectiveness of cost-sharing in a model with two competing manufacturer–retailer supply chains by offering substitutable products, and demonstrated that cooperative advertising improves consumer welfare by intensifying competition between supply chains. Karray (2015) used a game theoretic model to investigate equilibrium strategies for horizontal and vertical cooperative advertising in supply chains, revealing that in both centralized and decentralized channel competing settings, horizontal cooperative advertising among retailers influences vertical joint promotion. Chutani and Sethi (2018) modeled a Stackelberg differential game to explore dynamic cooperative advertising decisions, investigating the role of competition at both the manufacturer and retailer levels. Their findings indicate that high competition at the manufacturer level leads to optimal positive subsidy rates, while increased competition at the retailer level results in reduced support.

Our research focuses on advertising competition within SEA, which has been considered in both empirical studies and model analysis (Golden and Horton, 2021; Bhattacharya et al., 2022).

In the empirical studies, Animesh et al. (2011) examined the interaction between a firm's positioning strategy, ad rank in sponsored search listings, and advertising competitive intensity, finding that these factors significantly influence ad effectiveness and are critical strategic variables for firms in competitive markets. Yang et al. (2014) utilized a Bayesian estimation approach to analyze the effects of competition on keyword performance variables, and found that the number of competing advertisements significantly affects the baseline click volume, decay factor, and value per click. Based on a cross-sectional dataset from retailer firms, Ayanso and Karimi (2015) empirically showed that keyword competition significantly moderates the relationship between ad positioning and its determinants in the SEA market, with a more pronounced effect on multi-channel retailers than on web-only retailers.

In the model analysis, Xu et al. (2011) explored how to endogenously assess the valuation of superior advertising positions within a price competition framework, investigating the resulting outcomes in location competition and patterns of price dispersion using a game-theoretic model. They found that a prominent advertising position is not always advantageous for a firm with a competitive edge, as its profitability depends on balancing increased demand against the potential for higher prices when weaker competitors win the position. Based on a model of advertisers' rational competitive preference, Yuan et al. (2017) established an equilibrium solution, termed the upper bound Nash equilibrium, which effectively characterizes advertisers' competitive bidding behavior in sponsored search auctions. Chen and Guo (2022) developed a game-theoretic model to analyze price competition between a leading retailer and a smaller third-party seller offering identical products, finding that low-cost advertising through search engines significantly promotes platform openness and retail partnership. Tunuguntla et al. (2023) developed a multi-period, dynamic programming model that enables retailers to optimize their portfolio of generic and branded bids in a highly competitive SEA market.

## 2.3 Summary

Through our comprehensive review of the literature, two significant research gaps emerge. First, although cooperative advertising strategies have been extensively studied in traditional advertising channels, their application in the unique context of SEA remains largely unexplored. The distinctive characteristics of SEA, including its dynamic nature, varying competition intensity, and unique parameters (e.g., quality score), create fundamentally different challenges that cannot be adequately addressed by traditional cooperative advertising frameworks. Second, although advertising competition has been extensively studied both in cooperative advertising and SEA separately, its integration with cooperative advertising strategies in SEA remains under-researched. Given that competition is an inherent characteristic of SEA markets and significantly influences cooperative strategies' effectiveness, this limitation represents a critical gap in our understanding of how competition shapes cooperative advertising decisions in SEA.

In the existing literature, Cao and Ke (2019) is the only research that explored optimal cooperative advertising strategies in SEA while considering competition in position auction. They found that, unlike traditional media, manufacturers benefit from selective collaboration with only a subset of initially identical retailers in SEA. However, their approach employs a static model that may not adequately capture the dynamic characteristics of the SEA market. Moreover, their analysis is restricted to auction-level competition, overlooking market-level competition which is more prevalent in practice.

Table 1 compares the related literature across four dimensions: the inclusion of cooperative advertising strategies, use of dynamic models, consideration of advertising competition, and focus on the SEA market context. As illustrated in Table 1, to the best of our knowledge, we are the first to construct dynamic models for cooperative advertising strategies in SEA that more accurately reflect the evolving characteristics of the SEA market. Furthermore, we conduct equilibrium analysis in both scenarios with and without competition while broadening the scope of analysis by shifting the focus from auction-level competition to more extensive market-level competition. Our work addresses a significant gap in the existing literature and allows for a more nuanced understanding of how cooperative advertising strategies evolve under different competitive scenarios in the SEA market.

Table 1. Summary of some closely related works.

| Reference | Cooperative advertising strategies | Dynamic models | Advertising competition | SEA market |
|---|---|---|---|---|
| Animesh et al. (2011) | | | Yes | Yes |
| Ayanso and Karimi (2015) | | | Yes | Yes |
| Cao and Ke (2019) | Yes | | Yes | Yes |
| Chaab and Rasti-Barzoki (2016) | Yes | | | |
| Chen and Guo (2022) | | | Yes | Yes |
| Chutani and Sethi (2018) | Yes | Yes | Yes | |
| Ezimadu (2019) | Yes | Yes | | |
| He et al. (2009) | Yes | Yes | | |
| Hong et al. (2023) | Yes | | | |
| Jena et al. (2017) | Yes | | | |
| Karray (2015) | Yes | | Yes | |
| Karray et al. (2022) | Yes | Yes | | |
| Liu et al. (2014) | Yes | Yes | Yes | |
| Martín-Herrán and Sigué (2023) | Yes | Yes | | |
| Sarkar et al. (2020) | Yes | | | |
| Tunuguntla et al. (2023) | | | Yes | Yes |
| Xu et al. (2011) | | | Yes | Yes |
| Yang et al. (2014) | | | Yes | Yes |
| Yuan et al. (2017) | | | Yes | Yes |
| This study | Yes | Yes | Yes | Yes |

## 3. Model formulation

In this section, we establish cooperative advertising models in SEA, considering two scenarios illustrated in Figure 3. In Scenario I, the manufacturer adopts two strategies to promote products, involving direct-to-consumer advertising and collaboration with a retailer (Retailer-1). In a collaborative effort to incentivize the retail partner, the manufacturer provides a subsidy rate, denoted as $\theta$, which covers a portion of the retailer's advertising expenditure. In Scenario II, an additional independent retailer (Retailer-2) competes with Retailer-1. At time $t$, the advertising efforts are denoted as $u_1(t)$ for Retailer-1, $u_2(t)$ for Retailer-2 and $v(t)$ for the manufacturer. In the following, we formulate the advertising response functions for the two scenarios and the profit functions for both retailers and the manufacturer. The notations used in this paper are listed in Table 2.

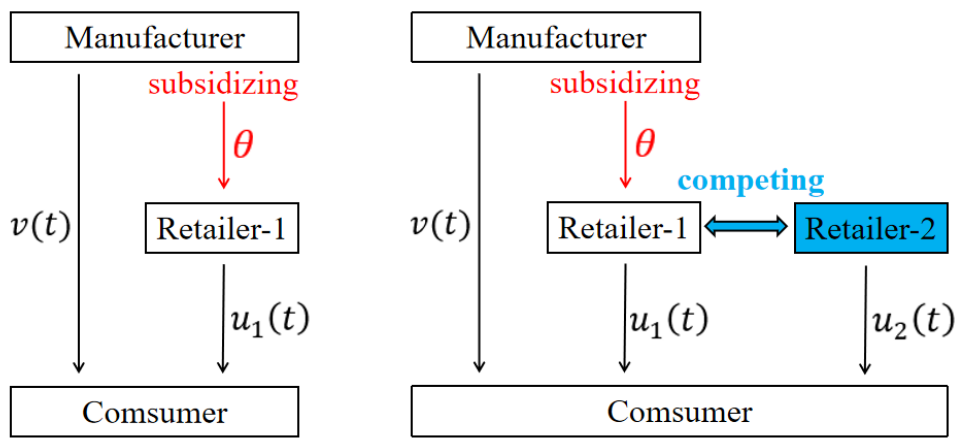

Scenario I                              Scenario II

Figure 3. Cooperative channels.

Table 2. List of notations.

| Symbol | Description |
|---|---|
| $t$ | Time $t$, $t \in [0, T]$ |
| $T$ | Terminal time |
| $i$ | Index of the retailer, $i = 1, 2$ |
| $M$ | Index of the manufacturer |
| $x_0$ | Initial market share |
| $r$ | Discount rate |
| $J$ | Profit functions |
| $V$ | Value functions |
| **Parameters** | |
| $\rho$ | Advertising effectiveness |
| $q(t)$ | Dynamic quality score at time $t$ |
| $c$ | Gross margin |
| **Decision Variables** | |
| $u_i(t)$ | Retailer $i$'s advertising effort at time $t$ |
| $v(t)$ | Manufacturer's advertising effort at time $t$ |
| $\theta$ | Manufacturer's subsidy rate for the retailer |
| **State Variable** | |
| $x(t)$ | Market share at time $t$ |

### 3.1 Advertising response function for SEA

Advertising response functions describe the relationship between advertising expenditures and market responses (e.g., market shares $x(t)$). One of the earliest and most important models was proposed by Vidale and Wolfe (1957), which explains how sales change when companies spend different amounts on advertising. Sethi (1983) improved upon the Vidale-Wolfe model with a square-root form of untapped market $\sqrt{1 - x(t)}$ to better capture desirable properties of advertising in the real world, including diminishing returns of advertising on market share, the existence of a saturation advertising level, and the word-of-mouth effect (Deal, 1979; Erickson, 1995; Krishnamoorthy et al., 2010; Yang et al., 2021). Specifically, Sethi's advertising response function is expressed as $\dot{x}(t) = \rho u \sqrt{1 - x(t)} - \delta x(t)$, where $\dot{x}(t)$ represents the rate of change of market shares $x(t)$ with respect to time $t$; $\rho u\sqrt{1-x(t)}$ explains how advertising effort $u$ positively influences the unsold portion of the market with a positive constant $\rho$; and decay rate $\delta$ measures how quickly market share decreases due to customers forgetting about the product or switching to competitors. While Sethi's model provides a strong foundation, we need to adapt it for the unique characteristics of SEA. Our first modification is to emphasize dynamic advertising effort $u(t)$, reflecting how SEA allows advertisers to adjust their campaigns in real time. Unlike traditional advertising, where campaigns are often fixed for long periods, SEA platforms enable advertisers to continuously modify their advertising content and strategies (Zhang et al., 2014; Dayanik and Sezer, 2023). Our second modification is the assumption of zero decay $\delta = 0$. Unlike traditional advertising where there's typically a time lag between ad exposure and purchase, during which sales may decay due to factors like forgetting or competition, SEA operates on a pay-per-click model where the realization of advertising cost (user click) and profit generation (also at click) occur simultaneously (Sethi, 1983; Yang et al., 2021). Additionally, SEA has very short advertising cycles (daily or even hourly) compared to longer product purchase cycles, further diminishing the impact of decay effects. Through simulation analysis, Yang et al. (2015a) validated that, compared to the one with zero decay, incorporating a small non-zero decay constant has minimal impact on market share, supporting the zero decay assumption in SEA. Moreover, since the decay rate partially captures competitive factors and our research focuses on comparing scenarios with and without retail competition, we set the decay rate to zero in our model construction. This allows us to completely eliminate competition-related factors when modeling Scenario I (Without Retail Competition), effectively depicting a competition-free environment. For Scenario II (With Retail Competition), we consider market competition in cooperative advertising through explicit modeling of a competitive retailer (Retailer-2). Our third modification is integrating advertiser's dynamic quality scores $q(t)$ into the advertising response function. As one of the most representative features of SEA, the quality score $q(t)$ plays a crucial role in an advertiser's ability to secure market shares (Feng et al., 2007; Zhang and Feng, 2011), necessitating its inclusion in the advertising response function (Yang et al., 2015a).

In our work, within Scenario I, the dynamic advertising response for the alliance of Retailer-1 and the manufacturer at time $t$ is given as

$$\dot{x}(t) = [\rho_1 q_1(t) u_1(t) + \rho_M q_M(t) v(t)]\sqrt{1-x(t)}, x(0) = x_0 \in [0,1], \tag{1}$$

where $x(t)$ is the market share captured by the alliance of Retailer-1 and the manufacturer at time $t$; $x_0$ is the initial level of market share (at time $t = 0$); $\rho_i$ and $q_i(t)$ ($i = 1,2,M$) represent the advertising effectiveness and advertising quality score at time $t$ for Retailer-1, Retailer-2 and manufacturer, respectively.

Contrastingly, as Scenario II delves into a duopolistic setting, and the presence of a competitor (Retailer-2) introduces additional dynamics to the advertising response, the dynamic advertising response for the alliance of Retailer-1 and the manufacturer at time $t$ is given as

$$\dot{x}(t) = \left(\rho_1 q_1(t) u_1(t) + \rho_M q_M(t) v(t)\right)\sqrt{1-x(t)} - \rho_2 q_2(t) u_2(t)\sqrt{x(t)}, \tag{2}$$

where the negative term reflects the loss of market share resulting from Retailer-2's competition.

### 3.2 Profit function of the retailers and the manufacturer

For the Retailer-1 and Retailer-2, we employ a standard discounted profit function $J_i$ ($i = 1,2$), which has been extensively studied in advertising literature (Feinberg, 2001; Reddy et al, 2016), given as

$$J_1 = \int_0^T [c_1 x(t) - (1-\theta) u_1(t)^2] e^{-rt} \, dt, \tag{3}$$

$$J_2 = \int_0^T [c_2 (1-x(t)) - u_2(t)^2] e^{-rt} \, dt, \tag{4}$$

where $c_i$ $(i = 1, 2)$ is the gross margin of Retailer-$i$ obtained from a unit of market share; $\theta$ is the constant subsidy rate from the manufacturer to Retailer-1; and $u_i(t)^2$ $(i = 1, 2)$ denote the advertising expenditures. The quadratic form reflects diminishing return of advertising effort $u_i(t)$, consistent with the law of diminishing marginal utility (Fruchter and Kalish, 1997; Ma et al, 2021). Thus, $c_1 x(t) - (1 - \theta) u_1(t)^2$ and $c_2 (1 - x(t)) - u_2(t)^2$ represent the profits obtained from advertising campaigns for Retailer-1 and Retailer-2 at time $t$, respectively. The term $e^{-rt} \in (0,1]$ serves as the discount factor, reducing future profits to their present value and accounting for the time value of money.

Similarly, for the manufacturer, the profit function can be described by

$$J_M = \int_0^T [c_M x(t) - v(t)^2 - \theta u_1(t)^2] e^{-rt} \, dt, \tag{5}$$

where $c_M$ is the gross margin of the manufacturer obtained from a unit of market share. Given that $\theta u_1(t)^2$ has been subsidized to Retailer-1, the profit obtained from advertising campaigns at time $t$ is $c_M x(t) - v(t)^2 - \theta u_1(t)^2$.

In summary, the dynamic cooperative advertising problems in SEA are formulated as differential games in two scenarios. Scenario I is defined by Equations (1), (3) and (5), with two players where $u_1(t)$ and $v(t)$ serve as control variables. Scenario II is defined by Equations (2), (3), (4) and (5) with three players, where $u_1(t)$, $u_2(t)$ and $v(t)$ are the control variables. The state variable in both scenarios is $x(t)$, and $\theta$ is considered as a decision variable of the manufacturer. These differential games can be solved using the Hamilton-Jacobi-Bellman (HJB) equation, an effective solution framework that has been widely adopted in optimal control problems (Prasad and Sethi, 2009; He et al., 2009; Ezimadu and Nwozo, 2017; Chutani and Sethi, 2018). The HJB equation is founded on Bellman's Principle of Optimality, which states that *"an optimal policy has the property that, whatever the initial state and initial decision are, the remaining decisions must*

*constitute an optimal policy with regard to the state resulting from the first decision*" (Bellman, 1966). This principle enables the decomposition of a complex decision problem into a sequence of smaller subproblems. The mathematical formalization of this principle yields equations that characterize optimal policies. Specifically, the HJB equation is expressed as follows:

$$-\frac{\partial V(x(t),t)}{\partial t} = max_{u(t) \in \mathcal{U}} \left\{ f(x(t), u(t), t) + \frac{\partial V(x(t),t)}{\partial x} g(x(t), u(t), t) \right\}, \quad (6)$$

where $x(t)$ denotes the state variable of the system. The value function $V(x(t), t)$ represents the optimal profit from any state $x(t)$ and time $t$ to the terminal period. The term $-\frac{\partial V(x(t),t)}{\partial t}$ captures how the value function changes with respect to time $t$, where the negative sign indicates the irreversibility of time $t$. The control variable $u(t) \in \mathcal{U}$ represents the decision variable, where $\mathcal{U}$ is the feasible control set. The function $f(x(t), u(t), t)$ represents the instantaneous return. The term $\frac{\partial V(x(t),t)}{\partial x}$ captures how the value function changes with respect to state $x$. The dynamic function $g(x(t), u(t), t)$ describes how the state evolves over time. By solving the HJB equation, we can obtain the optimal value function $V^*(x(t), t)$ and subsequently derive the optimal control policy $u^*(x(t), t)$. The detailed analysis of how to apply the HJB equation to solve the two differential games in Scenario I and Scenario II will be presented in Sections 4 and 5, respectively.

## 4. Scenario I: no competing retailer exists

In Scenario I, the situation is modeled as a Stackelberg differential game involving two channel members: the manufacturer, acting as the leader, and Retailer-1 as the follower. The sequence of strategic decisions unfolds as follows: first, the manufacturer announces the subsidy rate $\theta$ for Retailer-1 alongside its own advertising effort $v(t)$; in response, Retailer-1 determines its optimal advertising effort $u_1(t)$ by solving the optimization problems aimed at maximizing the present value of its profit stream over a finite horizon; then, sales are realized. Thus, Retailer-1's and the manufacturer's optimal control problems are given by

$$max_{u_1(t) \geq 0} \left\{ J_1 = \int_0^T [c_1 x(t) - (1-\theta) u_1(t)^2] e^{-rt} \, dt \right\}$$

s.t. $\dot{x}(t) = (\rho_1 q_1(t) u_1(t) + \rho_M q_M(t) v(t)) \sqrt{1 - x(t)}, \; x(0) = x_0 \in [0,1],$ (7)

and

$$max_{\substack{v(t) \geq 0, \\ \theta \in [0,1]}} \left\{ J_M = \int_0^T [c_M x(t) - v(t)^2 - \theta u_1(x|v,\theta)^2] e^{-rt} \, dt \right\}$$

s.t. $\dot{x}(t) = (\rho_1 q_1(t) u_1(x|v,\theta) + \rho_M q_M(t) v(t)) \sqrt{1 - x(t)}, \; x(0) = x_0 \in [0,1].$ (8)

### 4.1 Equilibrium analysis

In the equilibrium analysis, we detail a two-step resolution process. In the first step, we initially solve the sub-problem $(8)^{sub}$ (i.e., the control problem (8) with a subsidy rate fixed), to derive the manufacturer's optimal advertising effort in feedback form, expressed as $v^*(x,\theta), \theta \in [0,1)$. Following this, the feedback advertising effort of Retailer-1 can be expressed as $u_1^*(x|v^*(x,\theta))$. The policies $v^*(x)$, and $u_1^*(x)$ constitute a feedback Stackelberg equilibrium of the sub-problem $(7) - (8)^{sub}$, which is time-consistent. By integrating these policies into the state equation (1), we determine the market share process $x^*(t,\theta)$, $t \in [0,T]$, along with the corresponding decisions $v^*(t,\theta)$ and $u_1^*(t,\theta)$. Consequently, the control problem (8) simplifies to the following form:

$$max_{\theta \in [0,1]} \int_0^T [c_M x^*(t,\theta) - v^*(t,\theta)^2 - \theta u_1^*(t,\theta)^2] e^{-rt} \, dt. \tag{9}$$

In the second step, we solve the control problem (9) to determine the optimal subsidy rate $\theta^*$. Once $\theta^*$ is identified, the feedback advertising efforts for the manufacturer and Retailer-1 can be expressed as $v^*(t)$ and $u_1^*(t)$, respectively.

Following this, we will now detail the solving process of the Stackelberg differential game $(7) - (8)^{sub}$ and present the equilibrium solutions in the propositions. To facilitate this, we define the value functions $V_1(x,t)$ and $V_M(x,t)$ for Retailer-1 and the manufacturer at time $t$ when $x(t) = x$, respectively. We derive the Hamilton-Jacobi-Bellman equations:

$$0 = V_{1t} + \max_{u_1 \geq 0}\{e^{-rt}[c_1 x - (1-\theta)u_1^2] + V_{1x}\left((\rho_1 q_1(t) u_1(t) + \rho_M q_M(t) v(t))\sqrt{1-x}\right)\}, \quad (10)$$

$$0 = V_{Mt} + \max_{v \geq 0}\{e^{-rt}[c_M x - v^2 - \theta u_1^2] + V_{Mx}\left((\rho_1 q_1(t) u_1(t) + \rho_M q_M(t) v(t))\sqrt{1-x}\right)\}, \quad (11)$$

where $V_1(x,T) = 0$, $V_M(x,T) = 0$, $V_{1t} = \partial V_1/\partial t$, $V_{Mt} = \partial V_M/\partial t$, $V_{1x} = \partial V_1/\partial x$, $V_{Mx} = \partial V_M/\partial x$. The first-order conditions for maximum function in equations (10) and (11) yield the optimal advertising decisions $u_1^*(t)$ and $v^*(t)$. The results are provided in Proposition 1.

**Proposition 1.** *For a given subsidy rate $\theta$, the feedback Stackelberg equilibrium of the game $(7) - (8)^{sub}$ is given as follows*

(a) *The optimal advertising effort of Retailer-1 is given by*

$$u_1^*(x,\theta) = \frac{\rho_1 q_1(t) e^{rt} V_{1x} \sqrt{1-x}}{2(1-\theta)}. \quad (12)$$

(b) *The optimal advertising effort of the manufacturer is given by*

$$v^*(x,\theta) = \frac{\rho_M q_M(t) e^{rt} V_{Mx} \sqrt{1-x}}{2}. \quad (13)$$

(c) *The value functions $V_1(x,t)$ and $V_M(x,t)$ for Retailer-1 and the manufacturer satisfy the following two partial differential equations (HJ equations):*

$$V_{1t} + c_1 e^{-rt} x + \frac{(\rho_1 q_1(t))^2 e^{rt} V_{1x}^2 (1-x)}{4(1-\theta)} + \frac{(\rho_M q_M(t))^2 e^{rt} V_{1x} V_{Mx}(1-x)}{2} = 0, \quad (14)$$

$$V_{Mt} + c_M e^{-rt} x + \frac{(\rho_M q_M(t))^2 e^{rt} V_{Mx}^2 (1-x)}{4} - \frac{(\rho_1 q_1(t))^2 \theta e^{rt} V_{1x}^2 (1-x)}{4(1-\theta)^2} + \frac{(\rho_1 q_1(t))^2 e^{rt} V_{1x} V_{Mx}(1-x)}{2(1-\theta)} = 0. \quad (15)$$

**Proof.** See the Appendix A1.

We further define the value functions $V_1$ and $V_M$ to be linear with $x$, specified as

$$V_1(x,t) = e^{-rt}(\alpha_1(t) + \beta_1(t)x), \quad V_M(x,t) = e^{-rt}(\alpha_M(t) + \beta_M(t)x), \tag{16}$$

where $\alpha_1(t), \beta_1(t), \alpha_M(t), \beta_M(t)$ are time-dependent functions, $t \in [0,T]$. Since $V_1(x,T) = 0$ and $V_M(x,T) = 0$, by setting $t = T$ in equation (16), we have

$$V_1(x,T) = e^{-rT}(\alpha_1(T) + \beta_1(T)x) = 0, \tag{17}$$

$$V_M(x,T) = e^{-rT}(\alpha_M(T) + \beta_M(T)x) = 0, \tag{18}$$

where the terminal value conditions $\alpha_1(T) = 0$, $\beta_1(T) = 0$, $\alpha_M(T) = 0$ and $\beta_M(T) = 0$ are required. Substituting equation (16) into equations (14) and (15), and observing the time-variant coefficients of $x$ on both sides of the two equations, we derive four simultaneous ordinary differential equations that can be solved to obtain the four time-variant coefficients $\alpha_1(t), \beta_1(t), \alpha_M(t), \beta_M(t)$, which are summarized in Proposition 2.

**Proposition 2.** (a) *The functions $\alpha_1(t)$, $\beta_1(t)$ in the value function of Retailer-1 $V_1(x,t)$ satisfy the following equations*

$$\dot{\alpha}_1 = r\alpha_1 - \frac{(\rho_1 q_1(t))^2 \beta_1^2}{4(1-\theta)} - \frac{(\rho_M q_M(t))^2 \beta_1 \beta_M}{2}, \quad \alpha_1(T) = 0, \tag{19}$$

$$\dot{\beta}_1 = r\beta_1 + \frac{(\rho_1 q_1(t))^2 \beta_1^2}{4(1-\theta)} + \frac{(\rho_M q_M(t))^2 \beta_1 \beta_M}{2} - c_1, \quad \beta_1(T) = 0. \tag{20}$$

(b) *The functions $\alpha_M(t), \beta_M(t)$ in the value function of the manufacturer $V_M(x,t)$ satisfy the following equations*

$$\dot{\alpha}_M = r\alpha_M - \frac{(\rho_1 q_1(t))^2 \beta_M^2}{4} + \frac{(\rho_1 q_1(t))^2 \theta \beta_1^2}{4(1-\theta)^2} - \frac{(\rho_1 q_1(t))^2 \beta_1 \beta_M}{2(1-\theta)}, \quad \alpha_M(T) = 0, \tag{21}$$

$$\dot{\beta}_M = r\beta_M + \frac{(\rho_M q_M(t))^2 \beta_M^2}{4} - \frac{(\rho_1 q_1(t))^2 \theta \beta_1^2}{4(1-\theta)^2} + \frac{(\rho_1 q_1(t))^2 \beta_1 \beta_M}{2(1-\theta)} - c_M, \quad \beta_M(T) = 0. \tag{22}$$

(c) *The optimal advertising efforts of Retailer-1 and the manufacturer are given by*

$$u_1^* = \frac{\rho_1 q_1(t) \beta_1(t) \sqrt{1-x}}{2(1-\theta)}, \quad v^* = \frac{\rho_M q_M(t) \beta_M(t) \sqrt{1-x}}{2}. \tag{23}$$

(d) *The state trajectory $x(t)$ corresponding to the equilibrium is the solution of the initial value problem*

$$\dot{x} = \left((\rho_1 q_1(t))^2 \frac{\beta_1(t)}{2(1-\theta)} + (\rho_M q_M(t))^2 \frac{\beta_M(t)}{2}\right)(1-x), \; x(0) = x_0. \tag{24}$$

**Proof.** See the Appendix A2.

We can obtain the optimal advertising efforts $(u_1^*(t,\theta), v^*(t,\theta))$ by simultaneously solving equations (20), (22) and (24), and then determine the optimal subsidy rate by maximizing the corresponding profit. We consider two cases. In the first case, the manufacturer aims to maximize its own profit, denoting the optimal subsidy rate as $\theta^*$ (i.e., the non-integrated subsidy rate), a concept widely discussed in the literature (Karray and Amin, 2015; Ma et al., 2021). In the second case, the manufacturer aims to maximize the profit of the entire supply chain, denoting the optimal subsidy rate as $\bar{\theta}$ (i.e., the integrated subsidy rate), a scenario commonly explored in studies focusing on channel coordination (Zhang et al., 2013; He et al., 2019). The objective functions for the two cases are given as follows

$$J_M(\theta) = \int_0^T [c_M x^*(t,\theta) - v^*(t,\theta)^2 - \theta u_1^*(t,\theta)^2] e^{-rt} \, dt, \; \theta \in [0,1), \tag{25}$$

$$J_{R_1+M}(\theta) = J_{R_1}(\theta) + J_M(\theta)$$

$$= \int_0^T [(c_1 + c_M)x^*(t,\theta) - u_1^*(t,\theta)^2 - v^*(t,\theta)^2] e^{-rt} \, dt, \; \theta \in [0,1). \tag{26}$$

We present a computational solution to derive concrete values for $\theta^*$ and $\bar{\theta}$. The solution process is given in Algorithm 1. Algorithm 1 includes one total procedure and one embedded function. The embedded function "RealtimeDecisions" determines the optimal advertising efforts of Retailer-1 and the manufacturer $(u_1^*(t,\theta), v^*(t,\theta))$. Given $u_1^*(t,\theta)$ and $v^*(t,\theta)$, we calculate the manufacturer's objective profit functions $J_M(\theta)$ and $J_{R_1+M}(\theta)$ according to

Equations (25) and (26) for each $\theta \in [0,1)$. Finally, we take the value of $\theta^*$ as $\theta$ that maximizes the profit $J_M(\theta)$ and $\bar{\theta}$ as $\theta$ that maximizes the profit $J_{R_1+M}(\theta)$ in the total procedure.

---

**Algorithm 1**. (Optimal subsidy rates for two cases)

---

**Input:** $\beta_1(t,\theta), \beta_M(t,\theta), x^*(t,\theta)$
**Output:** $\theta^*, \bar{\theta}$
**Procedure:**
    For $\theta \leftarrow [0,1)$ do
        For $t \leftarrow [0,T]$ do
            Observe $\beta_1(t,\theta), \beta_M(t,\theta), x^*(t,\theta)$.
            $(u_1^*(t,\theta), v^*(t,\theta)) \leftarrow RealtimeDecisions(\beta_1(t,\theta), \beta_M(t,\theta), x^*(t,\theta))$
        End for
        Calculate $J_M(\theta)$ and $J_{R_1+M}(\theta)$ according to Equations (25) and (26).
        $\theta^* \leftarrow arg\ max\ J_M(\theta)$
        $\bar{\theta} \leftarrow arg\ max\ J_{R_1+M}(\theta)$
    End for
End procedure
    **Function** RealtimeDecisions
        $(\beta_1(t,\theta), \beta_M(t,\theta), x^*(t,\theta)) \leftarrow$ Equations (20)(22)(24)
        $u_1^*(t,\theta) \leftarrow \frac{\rho_1 q_1(t)\beta_1(t)\sqrt{1-x}}{2(1-\theta)}$,
        $v^*(t,\theta) \leftarrow \frac{\rho_M q_M(t)\beta_M(t)\sqrt{1-x}}{2}$
        **Return** $(u_1^*, v^*)$
    **End function**

---

Once we have determined the values of $\theta^*$ and $\bar{\theta}$, we can derive the equilibrium solution of the Stackelberg differential game, as outlined in equations $(7) - (8)$, by substituting $\theta = \theta^*$ into (19)-(24) in Proposition 3.

**Proposition 3** *The Stackelberg equilibrium of the game* $(7) - (8)$ *is given as follows*

(a) *The optimal advertising effort of Retailer-1 is given by*

$$u_1^*(t) = \frac{\rho_1 q_1(t)\beta_1(t,\theta^*)\sqrt{1-x(t,\theta^*)}}{2(1-\theta^*)}. \tag{27}$$

(b) *The optimal advertising effort of the manufacturer is given by*

$$v^*(t) = \frac{\rho_M q_M(t)\beta_M(t,\theta^*)\sqrt{1-x(t,\theta^*)}}{2}. \tag{28}$$

(c) *The optimal subsidy rate of the manufacturer is given by*

$$\theta^* = arg\ max\ J_M(\theta). \tag{29}$$

As the above system of differential equations (20), (22) and (24) is unlikely to be resolved through an analytic approach, similar to most studies (He et al., 2011; Chutani and Sethi, 2018; Pnevmatikos et al., 2018), we employ numerical analysis to further illustrate these theoretical results in the subsequent subsection.

**4.2 Numerical analysis**

We conduct numerical experiments to evaluate our model and strategies, and make sensitivity analysis with respect to the quality score $q(t)$ and the gross margin $c$. Our experimental evaluation serves two main purposes. First, we aim to evaluate how the quality score impacts optimal advertising efforts for Retailer-1 and the manufacturer, the corresponding trajectory of market share, and the two optimal subsidy rates. In this context, we will consider two settings: one with a constant quality score and another with a dynamic quality score. Second, we explore the effect of gross margin on the two optimal subsidy rates, and the profits of Retailer-1, the manufacturer and the cooperative channel, respectively.

*4.2.1 Impact of quality score q*

Our analysis first explores the impact of quality scores, examining both constant and dynamic quality scores of the manufacturer.

i. *Constant quality score $q_M$.* The values $q_M$ are given as 0.05, 0.1, 0.15, 0.2, 0.25 between $[0, T]$, respectively.

ii. *Dynamic quality score $q_M(t)$.* We consider two functional forms of the quality score $q_M(t)$.

- Linearly increasing between $[0, T]$, i.e., $q_M(t) = 0.05 + 0.2t/T$.
- Linearly decreasing between $[0, T]$, i.e., $q_M(t) = 0.25 - 0.2t/T$.

In each case, we further examine the influence of the manufacturer's quality score under two settings,

(a) Retailer-1 has low quality score ($q_1 = 0.05$),

(b) Retailer-1 has high quality score ($q_1 = 0.25$).

The base parameters are hypothesized as follows: $\rho_1 = \rho_M = 0.05, r = 0.05, c_1 = c_M = 200, T = 100, x_0 = 0.1$. Figure 4 illustrates the optimal advertising efforts of Retailer-1 and the manufacturer, as well as the market share trajectory of the Manufacturer-Retailer alliance, where parts (a) and (b) correspond to the situations with various constant values of $q_M$ for $q_1 = 0.05$ and $q_1 = 0.25$, respectively; part (c) and part (d) correspond to the situation with dynamic $q_M(t)$, specifically $q_M(t) = 0.05 + 0.2t/T$ in part (c) and $q_M(t) = 0.25 - 0.2t/T$ in part (d), with each applicable to both $q_1 = 0.05$ and $q_1 = 0.25$.

From Figure 4, we observe the following:

(1) Under each value of $q_M$, the optimal advertising efforts of both Retailer-1 and the manufacturer decrease over time. Correspondingly, their market share increases over time, exhibiting a decreasing marginal trend. This pattern is fully consistent with the common results derived from optimal control theory.

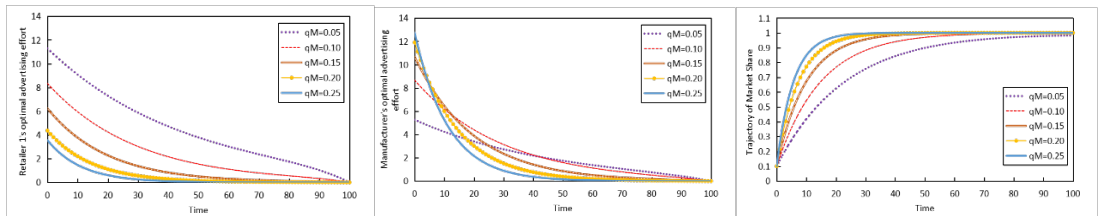

(a) $q_1 = 0.05$

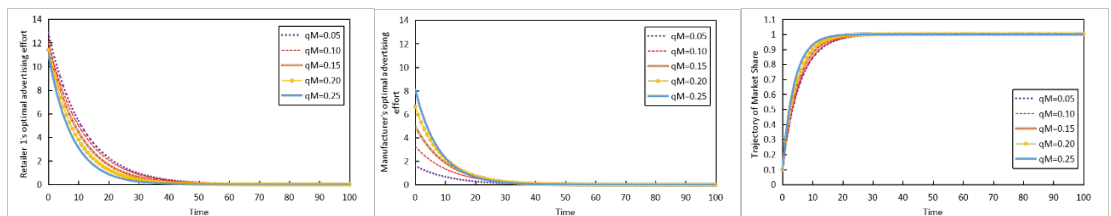

(b) $q_1 = 0.25$

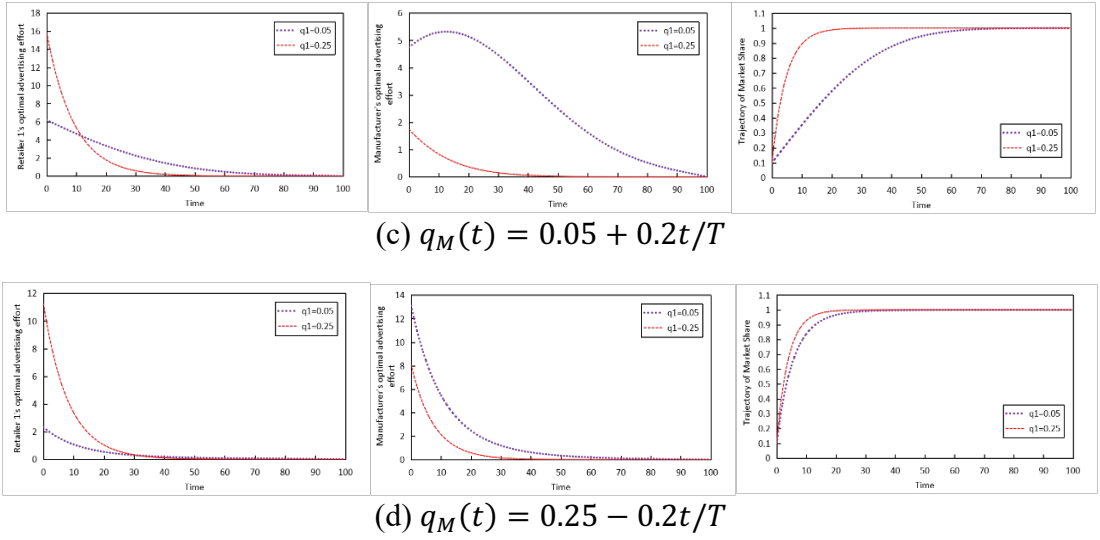

(c) $q_M(t) = 0.05 + 0.2t/T$

(d) $q_M(t) = 0.25 - 0.2t/T$

Figure 4. $u_1^*(t)$, $v^*(t)$ and $x^*(t)$ with different $q_M$.

(2) In the alliance between Retailer-1 and the manufacturer, the member with a higher quality score tends to invest more in advertising effort to secure a larger market share, while the other alliance member decreases the advertising effort, recognizing that a high market share can be maintained with the support of the ally. For example, Figures 4(a) and 4(b) illustrate that when the manufacturer's quality score $q_M$ is constant, an increase in $q_M$ stimulates the manufacturer to invest more in its advertising effort to capture more market share, while Retailer-1's advertising effort decrease. Similarly, Figures 4(c) and 4(d) show that the optimal advertising effort of Retailer-1 is higher than that of the manufacturer when $q_1 = 0.25$ ($q_1 \geq q_M(t), \forall t$), while the optimal advertising effort of the manufacturer is higher than Retailer-1 when $q_1 = 0.05$ ($q_1 \leq q_M(t), \forall t$).

Furthermore, to provide qualitative insight into how different forms of quality score impact the optimal advertising efforts of Retailer-1 and the manufacturer, as well as the corresponding market share, Figure 5 illustrates $u_1^*(t)$, $v^*(t)$, $x^*(t)$ under three forms of $q_M$ when $q_1 = 0.05$ and $q_1 = 0.25$: (a) the quality of the manufacturer's advertisement is kept to be a stable level, $q_M = 0.15$ (constant); (b) it improves through some efforts, $q_M(t) = 0.05 + 0.2t/T$ (dynamic-increasing); and (c) it declines due to some negligence, $q_M(t) = 0.25 - 0.2t/T$ (dynamic-decreasing). Note that the mean values of $q_M$ in all three forms are 0.15.

From Figure 5, it can be observed that under the three forms of the manufacturer's quality score:

(1) For Retailer-1, its optimal advertising efforts are highest when manufacturer's quality score shows an increasing trend, $q_M(t) = 0.05 + 0.2t/T$, in both cases $q_1 = 0.05$ and $q_1 = 0.25$. Such findings indicate that Retailer-1 is more inclined to increase its advertising effort when the manufacturer is committed to improving the quality of its advertisement.

(2) For the manufacturer, when observing that the Retailer-1's quality score is high ($q_1 = 0.25$) while its own quality is on a decline ($q_M(t) = 0.25 - 0.2t/T$), in order to secure a larger market share and foster sustained cooperation with Retailer-1, the manufacturer is compelled to augment its advertising effort significantly (i.e., provide the highest advertising efforts). Nonetheless, the strategy adjusts when the manufacturer perceives a low level of Retailer-1's advertising quality ($q_1 = 0.05$), alongside a deterioration in its own quality score. Initially, a higher advertising effort is employed to mitigate the adverse situation. However, if Retailer-1's quality score stays low and the manufacturer's quality score keeps decreasing, the manufacturer will promptly reduce its advertising effort to minimize loss.

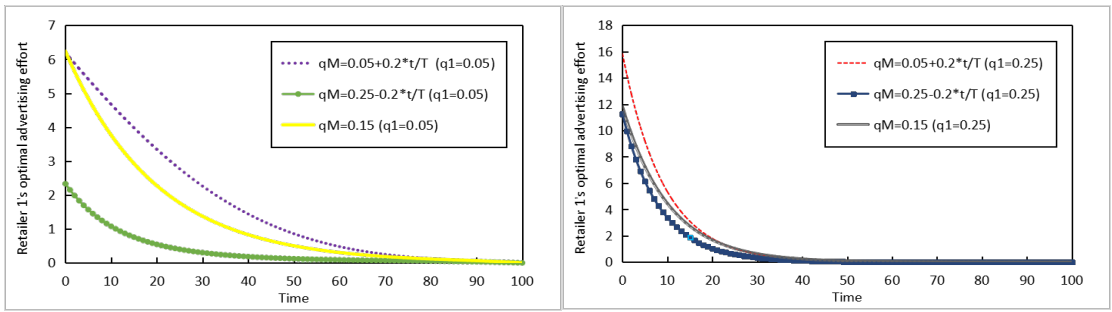

(a)

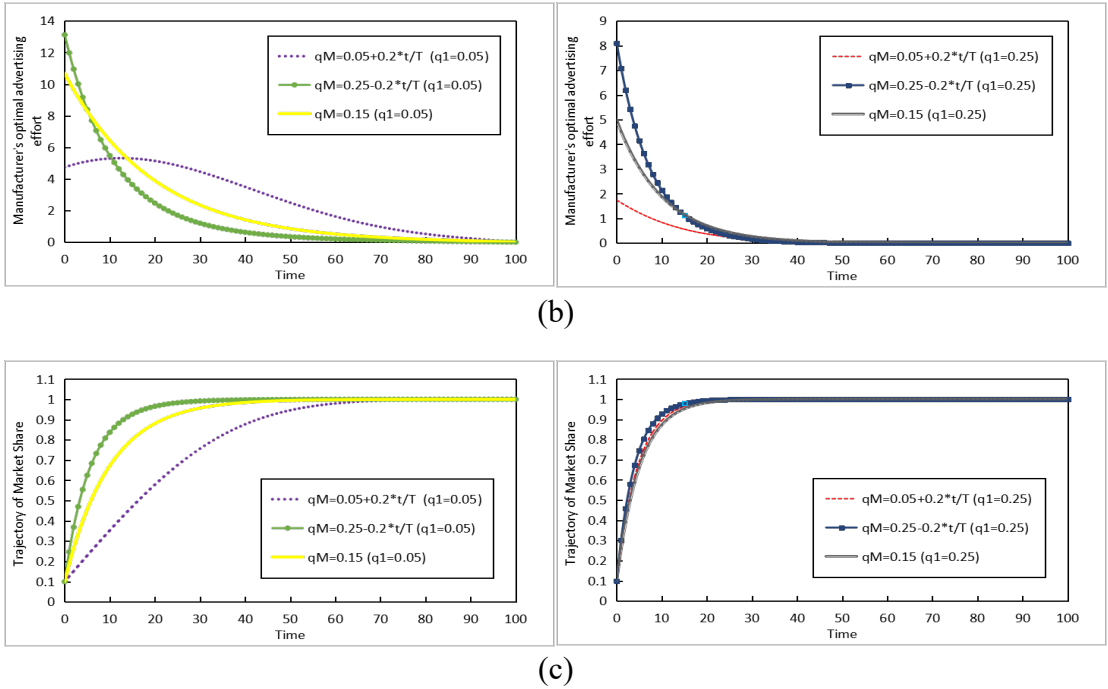

(b)

(c)

Figure 5. $u_1^*(t)$, $v^*(t)$ and $x^*(t)$ with different forms of $q_M(t)$.

Moreover, we examine the impact of quality score on the two optimal subsidy rates. Using a three-dimensional (3D) graph shown in Figure 6, we change $q_1$ and $q_M$ from 0.02 to 0.26 and show the observed results on subsidy rates, $\theta^*$ (i.e., the non-integrated subsidy rate) and $\bar{\theta}$ (i.e., the integrated subsidy rate).

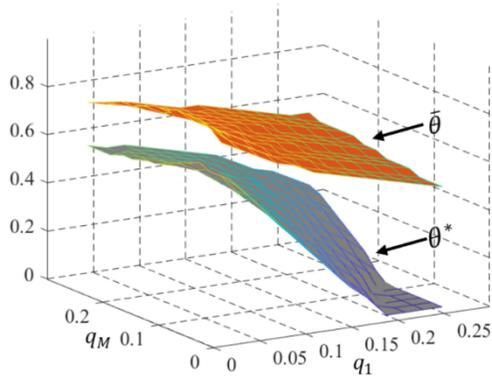

Figure 6. Two optimal subsidy rates $\theta^*$ and $\bar{\theta}$ with changes of $q_1$ and $q_M$.

From Figure 6, we observe the following:

(1) The optimal subsidy rates offered to Retailer-1 by the manufacturer exhibit a decrease with an increase in $q_1$, while they increase as $q_M$ rises. This pattern is attributed to strategic adjustments in subsidy allocations based on the perceived market share advantages and cost efficiencies derived from quality score variations. Specifically, the reduction in subsidy rates with higher $q_1$ value is rationalized by the manufacturer's recognition that Retailer-1 is capable of securing a larger market share without necessitating higher advertising expenditure, owing to the enhanced appeal of higher quality scores. Conversely, the increase in subsidy rates with rising $q_M$ value reflects the manufacturer's strategy to boost market share with lower advertising spend, leveraging higher quality's lower cost per click to reallocate budgets for more effective support of Retailer-1.

(2) The subsidy rate $\bar{\theta} > \theta^*$, and furthermore, $\bar{\theta}$ is less sensitive to variations in $q_1$ and $q_M$ compared to $\theta^*$. These observations imply that, compared to maximizing its own profit, more subsidy rates are required from the manufacturer to maximize the profit of the cooperative channel.

(3) Additionally, on the surface of $\theta^*$, a region identified as the $\theta^*$-Zero exists, characterized by a relatively large $q_1$ and a small $q_M$. Within the $\theta^*$-Zero region, the manufacturer provides no support to Retailer-1, indicating a non-cooperative equilibrium, where the manufacturer chooses to advertise directly to consumers rather than engaging in cooperative advertising by providing subsidy rate to Retailer-1.

### 4.2.2 Impact of gross margin c

We next analyze how gross margin impacts the profits of Retailer-1 and the manufacturer, the total profit of the cooperative channel and the two optimal subsidy rates. The base parameters $\{\rho_1, \rho_M, r, T, x_0\}$ remain consistent with those in section 3.2.1. Additionally, we set the quality score $q_1$ and $q_M$ to 0.15. Numerical results are presented in 3D graphics (Figure 7), where $c_1$ and $c_M$ vary in the interval $(0,300]$.

(1) In Figure 7(a), both Retailer-1's profit and the manufacturer's profit exhibit a direct correlation with their respective gross margins $c_1$ and $c_M$. Furthermore, the profit of the cooperative channel is shown to increase with enhancements in both $c_1$ and $c_M$. It is noteworthy, however, that there actually exists a positive effect of one member's gross margin on the other member's profit, which is not significantly shown in Figure 7(a), due to the dominant effect of each member's own gross margin.

(2) In Figure 7(b), as observed similarly in Figure 6, $\bar{\theta} > \theta^*$ and $\bar{\theta}$ is less sensitive to variations in $c_1$ and $c_M$ compared to $\theta^*$, and there exists a $\theta^*$-Zero region corresponding to gross margins $c_1$ and $c_M$. Interestingly, in some cases where Retailer-1's gross margin is quite small, the manufacturer can maximize its own profit and the cooperative channel at the same time, as indicated by the points where $\bar{\theta} = \theta^*$.

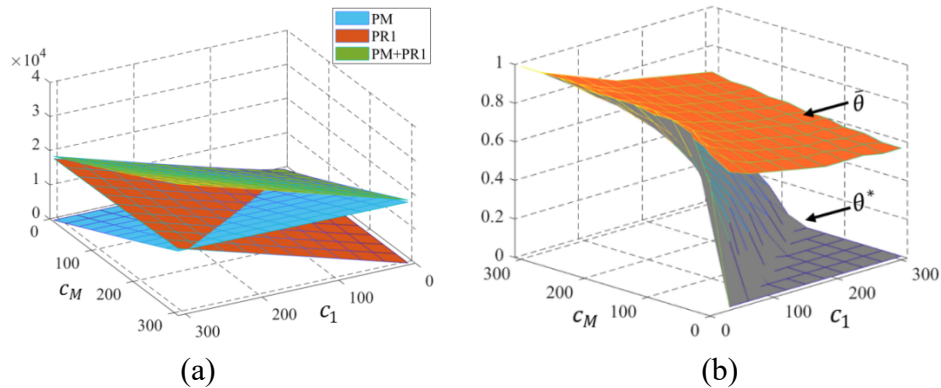

Figure 7. Impact of changes in $c_1$ and $c_M$ on profits of Retailer-1, the manufacturer and the cooperative channel, and two optimal subsidy rates $\theta^*$ and $\bar{\theta}$.

## 5. Scenario II: independent retailer competition

In Scenario II, besides Retailer-1, there is another retailer (Retailer-2) competing with

Retailer-1, who has not formed an alliance with the manufacturer. To model the interactions between a manufacturer and retailers within a supply chain framework, a Stackelberg differential game can be formulated, with the manufacturer acting as a leader and the retailers as followers. Additionally, a Nash differential subgame can be formulated to examine the competition between the retailers as they simultaneously determine their advertising efforts. This dual-game structure enables a nuanced analysis of both vertical and horizontal interactions within the supply chain. Specifically, the manufacturer announces the subsidy rate $\theta$ and its own advertising effort $v(t)$; in response, both retailers determine their advertising efforts $u_i(t)$ ($i = 1, 2$) by solving their optimization problems. Thus, the retailers' and the manufacturer's optimal control problems are given by

$$max_{u_1(t) \geq 0} \left\{ J_1 = \int_0^T [c_1 x - (1 - \theta) u_1(t)^2] e^{-rt} \, dt \right\}$$

$$max_{u_2(t) \geq 0} \left\{ J_2 = \int_0^T [c_2 (1 - x) - u_2(t)^2] e^{-rt} \, dt \right\}$$

s.t. $\dot{x}(t) = (\rho_1 q_1(t) u_1(t) + \rho_M q_M(t) v(t)) \sqrt{1 - x(t)} - \rho_2 q_2(t) u_2(t) \sqrt{x(t)}$,

$x(0) = x_0 \in [0, 1]$, (30)

and

$$max_{\substack{v(t) \geq 0, \\ \theta \in [0,1]}} \left\{ J_M = \int_0^T [c_M x - v(t)^2 - \theta u_1(x|v, \theta)^2] e^{-rt} \, dt \right\}$$

s.t. $\dot{x}(t) = (\rho_1 q_1(t) u_1(x|v, \theta) + \rho_M q_M(t) v(t)) \sqrt{1 - x(t)} - \rho_2 q_2(t) u_2(x|v, \theta) \sqrt{x(t)}$,

$x(0) = x_0 \in [0, 1]$. (31)

**5.1 Equilibrium analysis**

As in Scenario I, we aim to obtain the optimal decisions for the three players in Scenario II in two steps. In the first step, we solve the sub-problem $(31)^{sub}$ (i.e., the control problem (31) with a fixed subsidy rate) to derive the manufacturer's optimal advertising effort in feedback form, expressed as $v^*(x,\theta)$, $\theta \in [0,1)$. Furthermore, the feedback advertising efforts of Retailer-1 and Retailer-2 can be expressed as $u_1^*(x|v^*(x,\theta))$ and $u_2^*(x|v^*(x,\theta))$, respectively. The policies $v^*(x)$, $u_1^*(x)$ and $u_2^*(x)$ constitute a feedback Stackelberg equilibrium for the sub-problem (30)-(31)$^{\text{sub}}$, which is time-consistent. By substituting these policies into the state equation (2), we obtain the market share process $x^*(t,\theta)$, $t \in [0,T]$, along with the corresponding decisions $v^*(t,\theta)$, $u_1^*(t,\theta)$ and $u_2^*(t,\theta)$. Consequently, the control problem (31) is reduced to the same form as (9). In the second step, we solve the control problem (9) to determine the optimal subsidy rate $\theta^*$. Upon determining $\theta^*$, the feedback advertising efforts can be expressed as $v^*(t)$, $u_1^*(t)$ and $u_2^*(t)$. As the solution process is identical to that described in Section 4, we have omitted the repetitive details in this section and summarized the results in Table 4 and Proposition 4.

Table 4. Some preliminary results.

| | **HJB Equations** |
|---|---|
| Retailer-1 | $0 = V_{1t} + \max_{u_1 \geq 0}\{e^{-rt}[c_1 x - (1-\theta)u_1^2] + V_{1x}\dot{x}\}$ |
| Retailer-2 | $0 = V_{2t} + \max_{u_2 \geq 0}\{e^{-rt}[c_2(1-x) - u_2^2] + V_{2x}\dot{x}\}$ |
| Manufacturer | $0 = V_{Mt} + \max_{v \geq 0}\{e^{-rt}[c_M x - v^2 - \theta u_1^2] + V_{Mx}\dot{x}\}$ |
| | **HJ Equations** |
| Retailer-1 | $V_{1t} + c_1 e^{-rt} x + \frac{(\rho_1 q_1(t))^2 e^{rt} V_{1x}^2 (1-x)}{4(1-\theta)} + \frac{(\rho_2 q_2(t))^2 e^{rt} V_{1x} V_{2x} x}{2} + \frac{(\rho_M q_M(t))^2 e^{rt} V_{1x} V_{Mx} (1-x)}{2} = 0$ |
| Retailer-2 | $V_{2t} + c_2 e^{-rt}(1-x) + \frac{(\rho_1 q_1(t))^2 e^{rt} V_{1x} V_{2x} (1-x)}{2(1-\theta)} + \frac{(\rho_2 q_2(t))^2 e^{rt} V_{2x}^2 x}{4} + \frac{(\rho_M q_M(t))^2 e^{rt} V_{2x} V_{Mx} (1-x)}{2} = 0$ |
| Manufacturer | $V_{Mt} + c_M e^{-rt} x + \frac{(\rho_1 q_1(t))^2 e^{rt} V_{1x} V_{Mx} (1-x)}{2(1-\theta)} + \frac{(\rho_2 q_2(t))^2 e^{rt} V_{2x} V_{Mx} x}{2} + \frac{(\rho_M q_M(t))^2 e^{rt} V_{Mx}^2 (1-x)}{4} - \frac{\theta(\rho_1 q_1(t))^2 e^{rt} V_{1x}^2 (1-x)}{4(1-\theta)^2} = 0$ |
| | **The value functions** |
| Retailer-$i$ | $V_i(x,t) = e^{-rt}(\alpha_i(t) + \beta_i(t)x)$, $V_i(x,T) = 0$ |
| Manufacturer | $V_M(x,t) = e^{-rt}(\alpha_M(t) + \beta_M(t)x)$, $V_M(x,T) = 0$ |

**Proposition 4.** *For a given subsidy rate, the feedback Stackelberg equilibrium of the game* $(30) - (31)^{sub}$ *is given as follows*

(a) *The optimal advertising efforts of Retailer-1 and Retailer-2 are given by*

$$u_1^* = \frac{\rho_1 q_1(t) \beta_1(t) \sqrt{1-x}}{2(1-\theta)}, u_2^* = -\frac{\rho_2 q_2(t) \beta_2(t) \sqrt{x}}{2}. \tag{32}$$

(b) *The optimal advertising effort of the manufacturer is given by*

$$v^* = \frac{\rho_M q_M(t) \beta_M(t) \sqrt{1-x}}{2}. \tag{33}$$

(c) *The functions $\alpha_1(t)$, $\beta_1(t)$, $\alpha_2(t)$, $\beta_2(t)$ in the value functions of Retailer-1 and Retailer-2 satisfy the following equations*

$$\dot{\alpha}_1 = r\alpha_1 - \frac{(\rho_1 q_1(t))^2 \beta_1^2}{4(1-\theta)} - \frac{(\rho_M q_M(t))^2 \beta_1 \beta_M}{2}, \alpha_1(T) = 0, \tag{34}$$

$$\dot{\beta}_1 = r\beta_1 + \frac{(\rho_1 q_1(t))^2 \beta_1^2}{4(1-\theta)} - \frac{(\rho_2 q_2(t))^2 \beta_1 \beta_2}{2} + \frac{(\rho_M q_M(t))^2 \beta_1 \beta_M}{2} - c_1, \beta_1(T) = 0, \tag{35}$$

$$\dot{\alpha}_2 = r\alpha_2 - \frac{(\rho_1 q_1(t))^2 \beta_1 \beta_2}{2(1-\theta)} - \frac{(\rho_M q_M(t))^2 \beta_1 \beta_M}{2} - c_2, \alpha_2(T) = 0, \tag{36}$$

$$\dot{\beta}_2 = r\beta_2 + \frac{(\rho_1 q_1(t))^2 \beta_1 \beta_2}{2(1-\theta)} + \frac{(\rho_M q_M(t))^2 \beta_2 \beta_M}{2} - \frac{(\rho_2 q_2(t))^2 \beta_2^2}{4} + c_2, \beta_2(T) = 0. \tag{37}$$

(d) *The functions $\alpha_M(t)$, $\beta_M(t)$ in the value function of the manufacturer satisfy the following equations*

$$\dot{\alpha}_M = r\alpha_M - \frac{(\rho_M q_M(t))^2 \beta_M^2}{4} + \frac{(\rho_1 q_1(t))^2 \theta \beta_1^2}{4(1-\theta)^2} - \frac{(\rho_1 q_1(t))^2 \beta_1 \beta_M}{2(1-\theta)}, \alpha_M(T) = 0, \tag{38}$$

$$\dot{\beta}_M = r\beta_M + \frac{(\rho_M q_M(t))^2 \beta_M^2}{4} - \frac{(\rho_1 q_1(t))^2 \theta \beta_1^2}{4(1-\theta)^2} + \frac{(\rho_1 q_1(t))^2 \beta_1 \beta_M}{2(1-\theta)} - \frac{(\rho_2 q_2(t))^2 \beta_2 \beta_M}{2} - c_M, \beta_M(T) = 0. \tag{39}$$

(e) *The state trajectory $x(t)$ corresponding to the equilibrium is the solution of the initial value problem*

$$\dot{x} = \left((\rho_1 q_1(t))^2 \frac{\beta_1(t)}{2(1-\theta)} + (\rho_M q_M(t))^2 \frac{\beta_M(t)}{2}\right)(1-x) + \frac{(\rho_2 q_2(t))^2 \beta_2(t) x}{2}, x(0) = x_0. \tag{40}$$

**Proof.** See the Appendix A3.

Similarly, we consider two cases and obtain the values of $\theta^*$ and $\bar{\theta}$ through minimal modifications with $u_1 \to (u_1, u_2)$, $\beta_1 \to (\beta_1, \beta_2)$, Equations (20), (22), (24) → Equations (35), (37), (39), (40) in Algorithm 1. Once we have the values of $\theta^*$ and $\bar{\theta}$, we can then derive the equilibrium solution of the Stackelberg differential game (30)-(31) by substituting $\theta = \theta^*$ in (32)-(40).

Clearly, the obtained theoretical solution specified by the equations (32) and (33) is directly decided by the system of equations (35), (37), (39) and (40) which is more complex than those in Scenario I (Equations (20), (22) and (24)). To study such complex models, we resort to in-depth numerical simulations in section 5.2.

## 5.2 Numerical analysis

In this subsection, we perform numerical analysis to explore the impact of key parameters (i.e., the quality score and the gross margin) on profits and optimal subsidy rates. Similar to Scenario I, we demonstrate the influence of one parameter (e.g., quality score) by varying it while fixing the other (e.g., the gross margin) and vice versa. The base parameter values are provided in Table 5. Furthermore, to understand how the initial market share of the competing retailer (Retailer-2) affects equilibrium outcomes, we examine three initial values $x_0 = (0.1, 0.5, 0.9)$. This analysis allows us to assess the potential effects of market share shifts.

Table 5. Base parameter values.

| Parameter | $\rho_s$ | $q_s$ | $c_s$ | $r$ | $T$ | $x_0$ | |
|---|---|---|---|---|---|---|---|
| Value | 0.05 | 0.15 | 200 | 0.05 | 100 | 0.1 | $s \in \{1,2,M\}$ |

### 5.2.1 Impact of quality score q

To analyze how the profit of each channel member and the two optimal subsidy rates are influenced by variations in quality scores $q_i, i \in \{1,2,M\}$, we hold all other parameters constant at the values listed in Table 5. Given that there are three quality scores, we fix one of them at a time to facilitate a 3D representation of the results. Specifically, we explore the impact of varying $q_1$ and $q_M$ within the interval (0, 0.26], while setting $q_2$ at three distinct values: 0.04, 0.14, and 0.24, which allows us to quantify certain properties and gain insights. In the following discussion, we refer to the profits of Retailer-1, Retailer-2, and the manufacturer as $PR_1$, $PR_2$ and $PM$, respectively. The findings are illustrated in Figure 8.

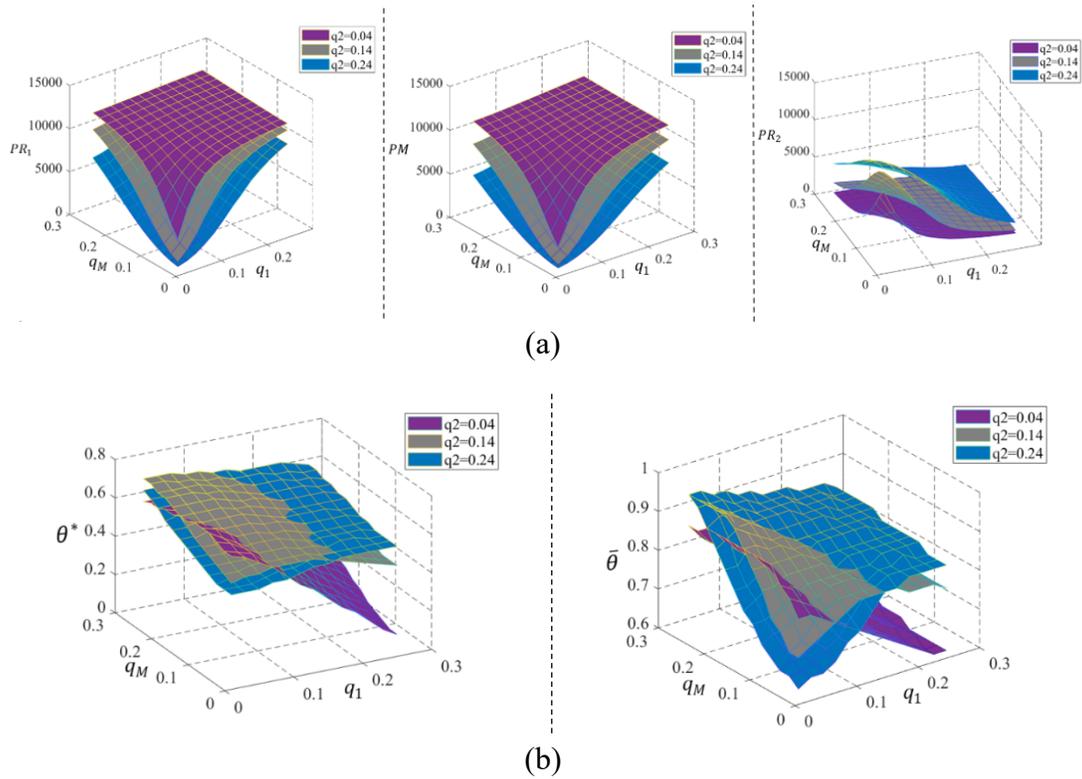

Figure 8. Profits of Retailer-1, the manufacturer and Retailer-2, and two optimal subsidy rates $\theta^*$ and $\bar{\theta}$ with changes of $q_1$ and $q_M$ when $q_2 = 0.04$, $q_2 = 0.14$ and $q_2 = 0.24$.

(1) Figure 8(a) illustrates that as the quality scores of Retailer-1 and the manufacturer ($q_1$ and $q_M$) increase, their profits rise, while Retailer-2's profit decreases. Conversely, an increase in Retailer-2's quality score ($q_2$) leads to an increase in its profit, while the profits of Retailer-1 and the manufacturer decline. This pattern demonstrates that higher quality score benefits the entity's profit and negatively impacts its competitor's profit.

(2) Figure 8(b) reveals that when $q_2$ is relatively small ($q_2 = 0.04$), similar to Scenario I, an increase in Retailer-1's quality score ($q_1$) leads to a significant decrease in the two optimal subsidy rates, while an increase in the manufacturer's quality score ($q_M$) results in a significant increase in these rates. Conversely, when $q_2$ is relatively large ($q_2 = 0.14$ and $q_2 = 0.24$), both $\theta^*$ and $\bar{\theta}$ are higher in most regions and are less influenced by changes in $q_1$ and $q_M$. Notably, an inverse effect of $q_1$ emerges beyond a certain threshold. These findings suggest that in face of heightened competition signaled by a high quality score of Retailer-2, the manufacturer tends to increase support for Retailer-1, largely independent of its own or Retailer-1's quality scores. This tendency disappears at lower values of $q_1$ and $q_M$ and is replaced by negative effect of $q_2$. In addition, there is no $\theta^*$-Zero region in the presence of Retailer-2's competition.

### 5.2.2 Impact of gross margin c

In the subsequent analysis, we examine the effects of gross margins $c_i, i \in \{1, 2, M\}$ on the profits of Retailer-1 ($PR_1$), Retailer-2 ($PR_2$), the manufacturer ($PM$), and the two optimal subsidy rates ($\theta^*$ and $\bar{\theta}$). For this purpose, we maintain the gross margin of Retailer-2 ($c_2$) at three distinct levels: $c_2 = 40$, $c_2 = 160$, and $c_2 = 280$, while allowing the gross margins of Retailer-1 ($c_1$) and the manufacturer ($c_M$) to vary within the interval $(0, 300]$. The outcomes of these variations are illustrated in Figure 9.

(1) In Figure 9(a), we observe that the profits of cooperative members (Retailer-1 and the manufacturer) increase with their gross margins ($c_1$ and $c_M$) and decrease with the higher gross margin of their competitor Retailer-2 ($c_2$). Similarly, Retailer-2's profit increases with its own gross margin ($c_2$) but is negatively affected by the gross margins of its competitors ($c_1$ and $c_M$). Notably, compared to Scenario I, the positive impact of changes in the manufacturer's gross margin on Retailer-1's profit is more pronounced, highlighting the significant role of gross margin in the cooperative alliance within the competitive retail environment.

(2) Figure 9(b) demonstrates that the gross margins of Retailer-1 ($c_1$) and the manufacturer ($c_M$) consistently influence the optimal subsidy rates $\theta^*$ and $\bar{\theta}$ in both Scenario I and Scenario II. It also shows a general trend where higher gross margin for Retailer-2 ($c_2$) increases these rates in most cases, but decreases them when $c_1$ and $c_M$ are low.

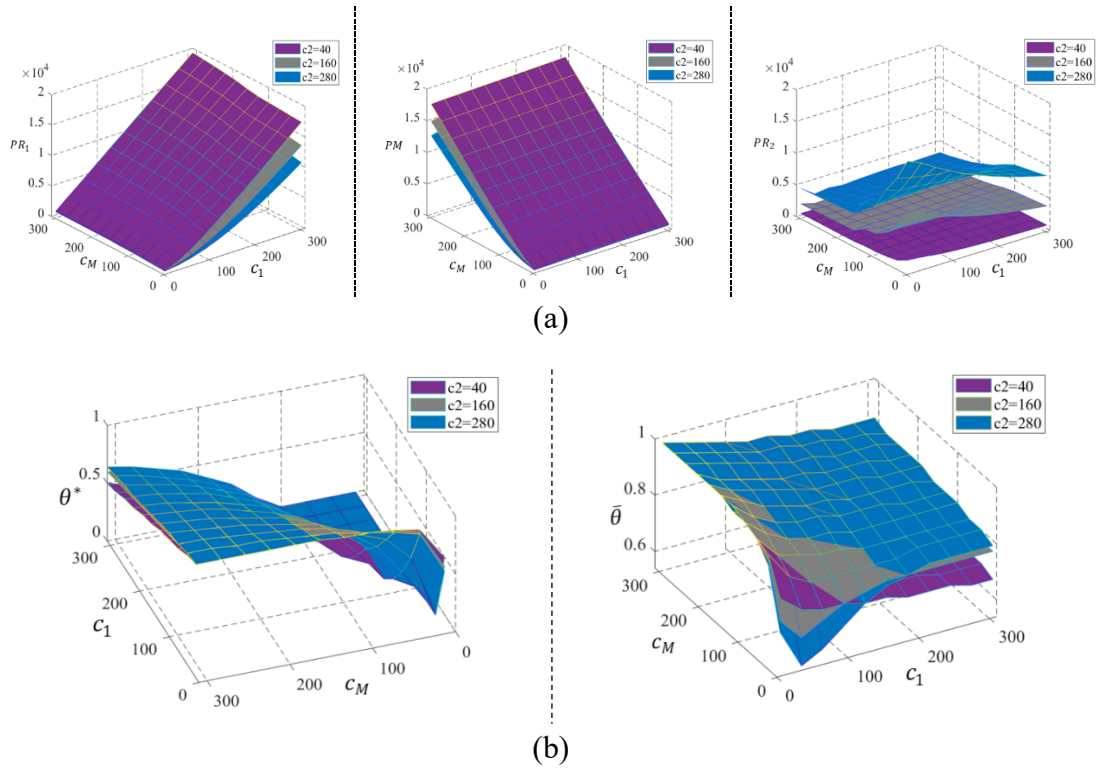

Figure 9. Profits of Retailer-1, the manufacturer and Retailer-2, and two optimal subsidy rates $\theta^*$ and $\bar{\theta}$ with changes of $c_1$ and $c_M$ when $c_2 = 40$, $c_2 = 160$ and $c_2 = 280$.

### 5.2.3 Impact of Retailer-2's initial market share

In this subsection, we focus on the outcomes associated with different initial market share. We consider three values of $x_0 = \{0.1, 0.5, 0.9\}$, where $x_0 = 0.1$ indicates that Retailer-2 having a higher initial market share $(1 - x_0 = 0.9)$, becomes the dominant entity in the market and a "Large competitor" to Retailer-1; $x_0 = 0.5$ indicates that Retailer-2 and Retailer-1 have identical initial market shares, positioning Retailer-2 as an "Equal competitor" to Retailer-1; $x_0 = 0.9$ signifies that Retailer-1 assumes the role of the dominant player, relegating Retailer-2 to the status of a "Small competitor". The three cases are summarized in Table 6.

Table 6. Retailers' three initial market shares.

| Initial market share | Large competitor | Equal competitor | Small competitor |
|---|---|---|---|
| $x_0$ (Retailer-1) | 0.1 | 0.5 | 0.9 |
| $1 - x_0$ (Retailer-2) | 0.9 | 0.5 | 0.1 |

Figures 10–12 display representative results, from which we derive the following key observations:

When Retailer-2 is categorized either as a "Large competitor" or an "Equal competitor", we observe that both Retailer-1 and the manufacturer initially exert high advertising efforts. These efforts gradually decrease at a decelerating rate before stabilizing, and ultimately, they diminish significantly. This trend is characterized as a Decrease Pattern. Concurrently, due to Decrease Pattern of advertising efforts of both Retailer-1 and the manufacturer, their market share starts from a lower level and increases over time. Conversely, Retailer-2's advertising effort exhibits an Increase Pattern, demonstrating a trajectory opposite to that of Retailer-1 and the manufacturer. Conversely, when Retailer-2 is considered a "Small competitor", the advertising efforts of Retailer-1 and the manufacturer follow an Increase Pattern, whereas Retailer-2's efforts align with a Decrease Pattern. This inversion also extends to the pattern of market share evolution. This observation suggests that initial market share inversely influences its own and its rivals' advertising efforts, with a higher market share correlating to increased efforts from competitors and reduced efforts from itself. Moreover, Retailer-2's initial market share negatively impacts the subsidy rates, which suggests that the manufacturer will provide more advertising support when it possesses a higher initial market share.

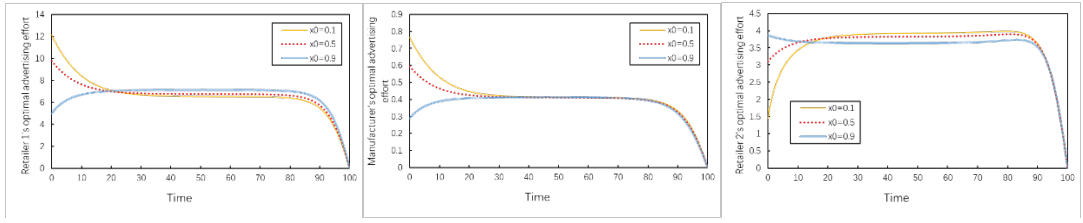

Figure 10. $u_1^*(t)$, $v^*(t)$ and $u_2^*(t)$ with different $x_0$.

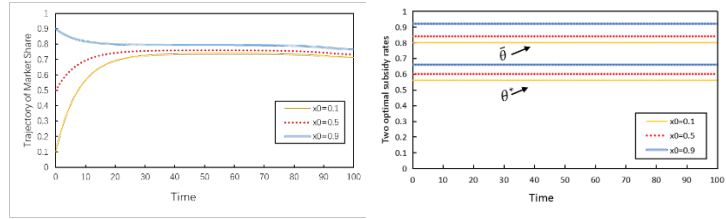

Figure 11. $x^*(t)$, $\theta^*$ and $\bar{\theta}$ with different $x_0$.

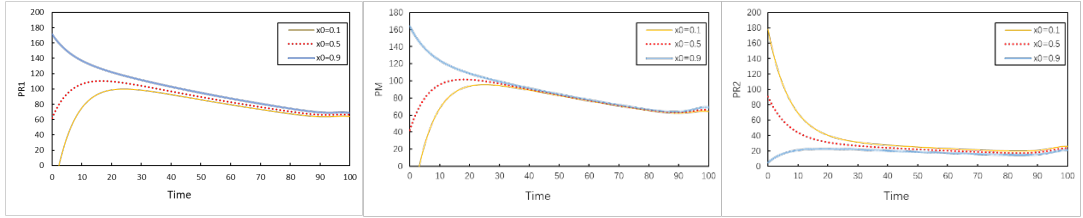

Figure 12. $PR_1$, $PM$ and $PR_2$ with different $x_0$.

## 6. The impact of retail competition

This section examines the impact of retail competition by comparing Scenario I (without retail competition, WORC) and Scenario II (with retail competition, WRC). Specifically, we analyze how retail competition influences both the optimal strategies of the Manufacturer-Retailer alliance (i.e., advertising efforts and subsidy rates) and channel performance (i.e., market share and profit). Furthermore, to understand the moderating role of quality scores in retail competition, we conduct our analysis under two distinct settings: (1) manufacturer's quality score variation, where we fix Retailer-1's quality score $q_1$ at either 0.05 (low level) or 0.25 (high level), while varying the manufacturer's quality score $q_M$ as 0.05, 0.1, 0.15, 0.2, 0.25; (2) retailer's quality score variation, where we fix manufacturer's quality score $q_M$ at either 0.05 (low level) or 0.25 (high level), while varying Retailer-1's quality score $q_1$ as $0.05, 0.1, 0.15, 0.2, 0.25$. The default values of parameters $(\rho_s, c_s, s \in \{1,2,M\}, T, r, x_0)$ remain consistent with those presented in Table 5.

### 6.1 Impact of retail competition on optimal strategies

We first explore the impact of retail competition on the trajectories of optimal advertising efforts for both Retailer-1 and the manufacturer (Figure 13), and the manufacturer's non-integrated and integrated subsidy rates (Figure 14).

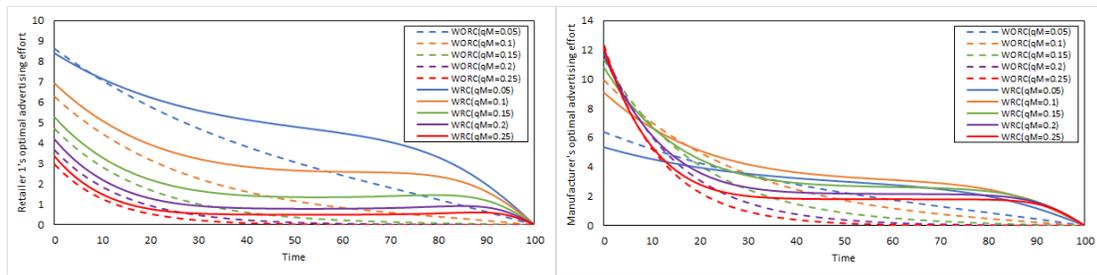

(a) $q_1 = 0.05$

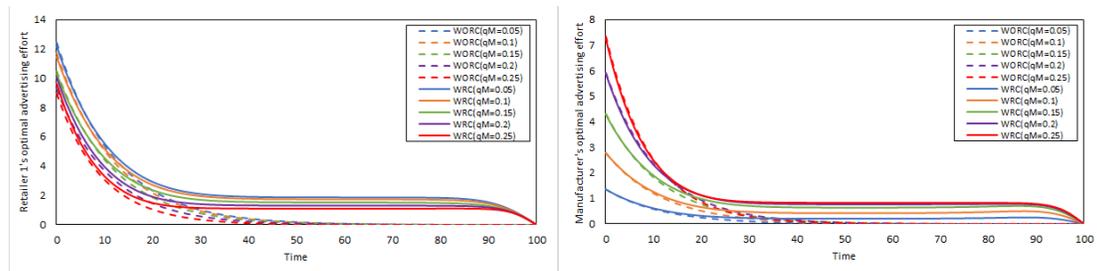

(b) $q_1 = 0.25$

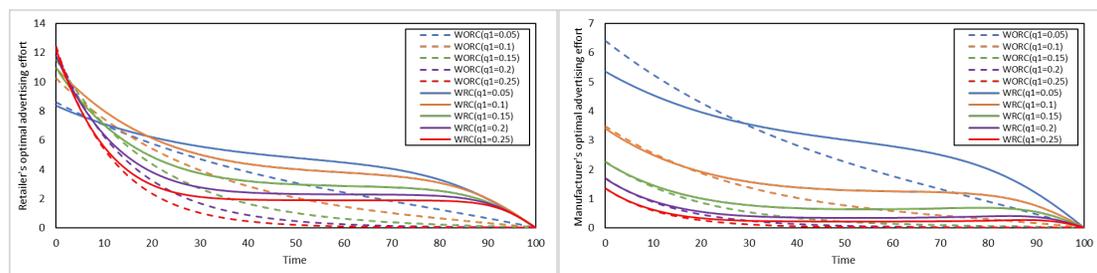

(c) $q_M = 0.05$

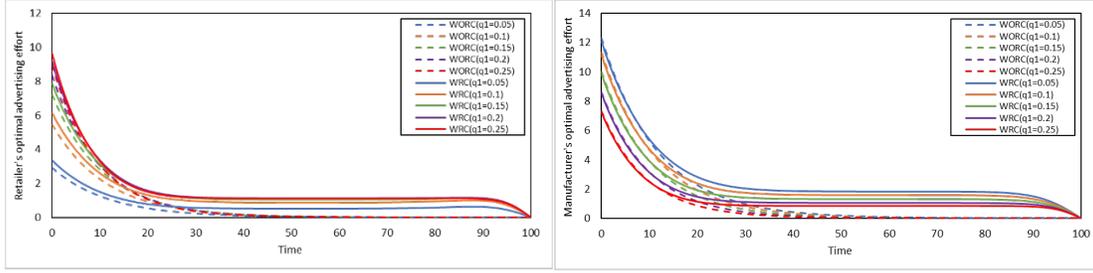

(d) $q_M = 0.25$

Figure 13. Impact of retail competition on the optimal advertising efforts of Retailer-1 and the manufacturer $(u_1^*(t), v^*(t))$.

Figure 13 shows that retail competition generally drives the Manufacturer-Retailer alliance (consisting of the manufacturer and Retailer-1) to increase their optimal advertising efforts. This effect is particularly significant when either member has a low quality score ($q_M = 0.05$ or $q_1 = 0.05$) in SEA. For any member with a low quality score, when its partner's quality score increases, the impact of retail competition on this member's optimal advertising effort gradually diminishes. However, for the partner with increasing quality score, the impact of retail competition on its own advertising effort remains relatively stable. As illustrated in Figure 13(a), when the Retailer-1's quality score is low ($q_1 = 0.05$), an increase in the manufacturer's quality score ($q_M: 0.05 \to 0.25$) leads to a gradual reduction in the impact of retail competition on Retailer-1's advertising effort (reflected by the narrowing gap between WRC and WORC scenarios). Meanwhile, the effect on the manufacturer's advertising effort remains relatively stable. This phenomenon suggests a quality score based complementarity effect in the alliance: when a member has a low quality score, the improvement in its partner's quality score can partially compensate for its competitive disadvantage, thus reducing its sensitivity to retail competition in terms of advertising effort decisions. For the partner with higher quality score, its optimal advertising strategy is primarily driven by its own quality score advantage, leading to a relatively stable response to retail competition. The same pattern is observed in Figure 13(c) when the manufacturer's quality score is low ($q_M = 0.05$). Figures 13(b) and 13(d) show that when either member in the alliance has a high quality score, the impact of retail competition on both members' advertising efforts remains stable regardless of changes in their partner's quality score. This indicates that a high quality score helps stabilize the alliance's advertising strategies under retail competition.

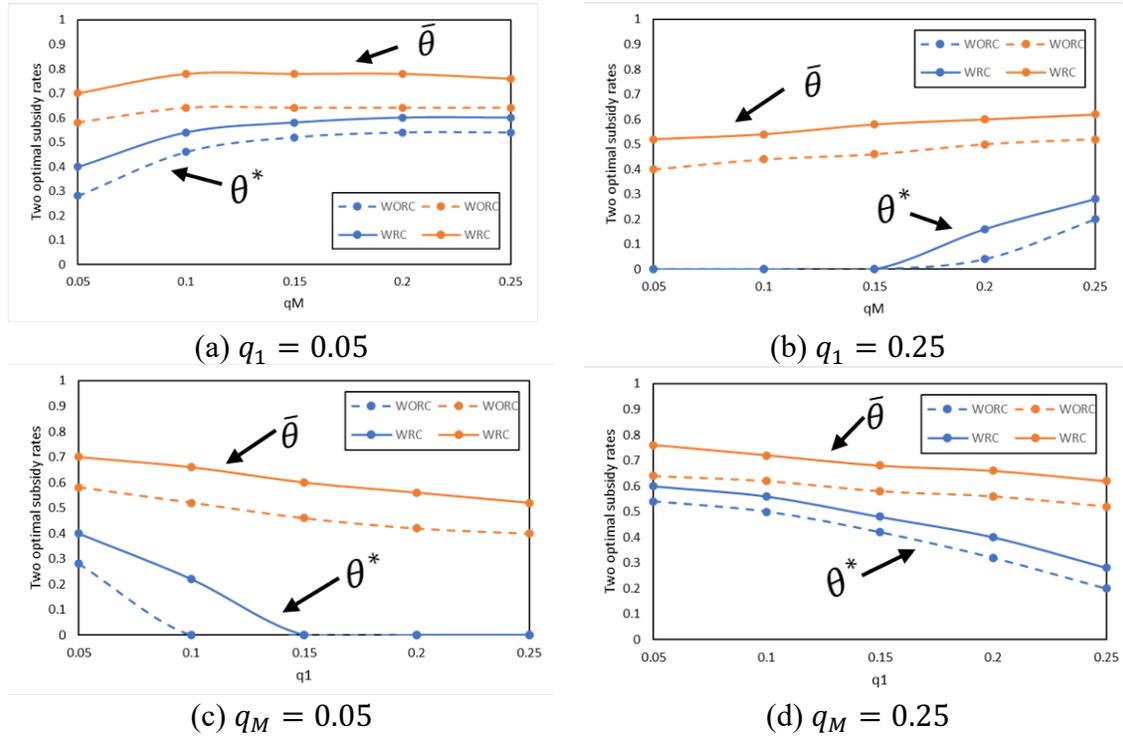

Figure 14. Impact of retail competition on manufacturer's optimal subsidy rates ($\theta^*, \bar{\theta}$).

Figure 14 illustrates how retail competition impacts the manufacturer's optimal subsidy rate decisions, including non-integrated subsidy rates $\theta^*$ and integrated subsidy rates $\bar{\theta}$. The results reveal that subsidy rates under retail competition are no lower than those in non-competitive scenarios. The impact of retail competition on $\bar{\theta}$ (integrated subsidy rate) remains relatively stable across different quality score combinations. However, retail competition's impact on $\theta^*$ (i.e., non-integrated subsidy rate) exhibits more complex patterns. Specifically, as illustrated in Figures 14(a) and 14(d), when the retailer's quality score is low or the manufacturer's quality score is high ($q_1 =$

0.05 or $q_M = 0.25$), the impact of retail competition on $\theta^*$ remains stable regardless of changes in its partner's quality score. In Figure 14(c), when the manufacturer's quality score is low ($q_M = 0.05$), the manufacturer stops providing subsidies when Retailer-1's quality score reaches 0.15 under the WRC scenario, while this happens when Retailer-1's quality score reaches 0.1 under the WORC scenario. This difference arises because under the WRC scenario, the cooperative alliance needs to build sufficient competitive advantage, requiring a higher quality score before manufacturer withdraws subsidies to retailer. In contrast, under the WORC scenario, the absence of competitive pressure allows the manufacturer to stop providing subsidies at a lower quality score of Retailer-1. In Figure 14(b), when Retailer-1's quality score is high ($q_1 = 0.25$), the manufacturer starts providing subsidies when the manufacturer's quality score reaches 0.15 under both the WRC and WORC scenarios. This identical behavior across scenarios occurs probably because when both the manufacturer and Retailer-1 have high quality scores, the manufacturer prioritizes overall supply chain quality balance, with retail competition having minimal impact on its subsidy behavior.

## 6.2 Impact of retail competition on channel performance

We investigate the impact of retail competition on the optimal trajectory of the market share (Figure 15), the individual profits of cooperative channel members (Figure 16) and the total channel profit (Figure 17).

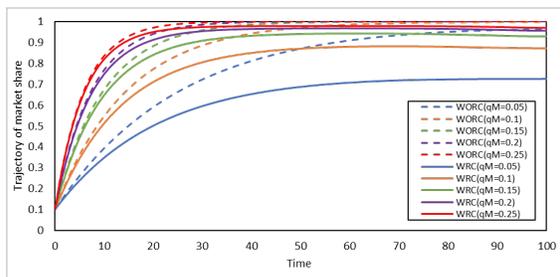

(a) $q_1 = 0.05$

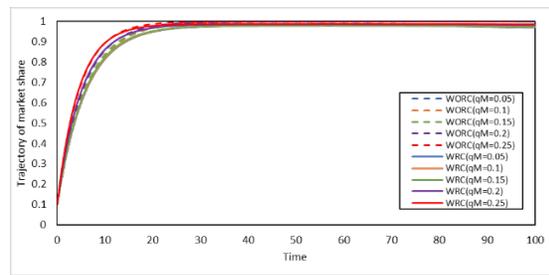

(b) $q_1 = 0.25$

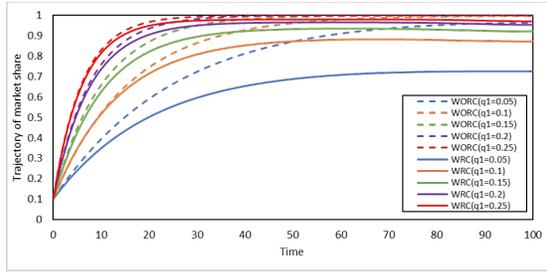
(c) $q_M = 0.05$

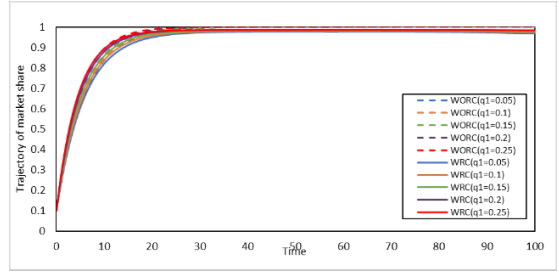
(d) $q_M = 0.25$

Figure 15. Impact of retail competition on the optimal trajectory of market share ($x^*(t)$).

Figure 15 reveals that retail competition reduces the optimal market share of the cooperative alliance. As illustrated in Figures 15(a) and 15(c), when either member in the Manufacturer-Retailer alliance has a low quality score ($q_1 = 0.05$ or $q_M = 0.05$), the negative impact of retail competition on market share diminishes as its partner's quality score increases. Figures 15(b) and 15(d) show that when either member has a high quality score ($q_1 = 0.25$ or $q_M = 0.25$), the impact of competition on the alliance's market share becomes less sensitive to changes in the partner's quality score. This finding suggests that in a retail competitive scenario, members with high quality score can help mitigate the negative impact of their partner's low quality scores on market share.

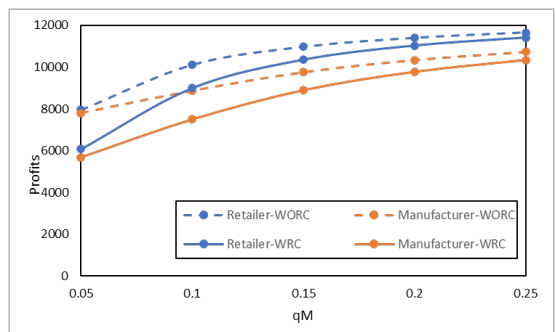
(a) $q_1 = 0.05$

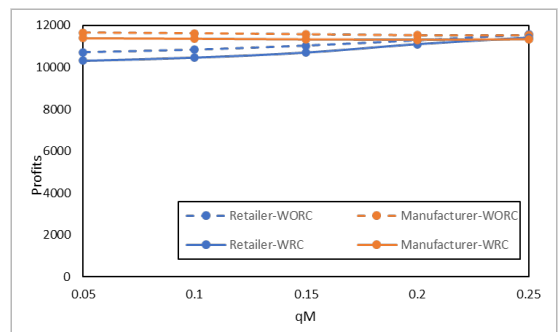
(b) $q_1 = 0.25$

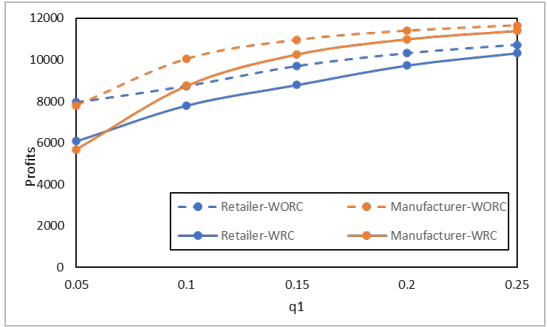
(c) $q_M = 0.05$

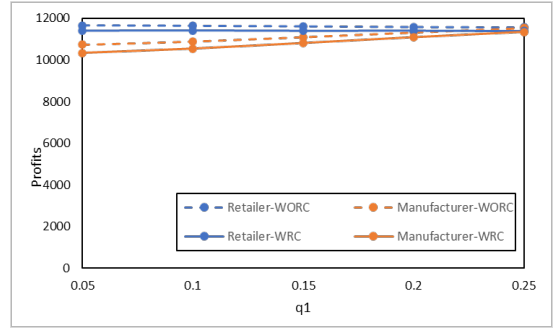
(d) $q_M = 0.25$

Figure 16. Impact of retail competition on profits of Retailer-1 and the manufacturer.

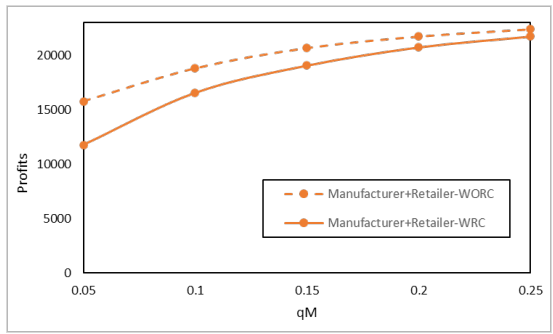
(a) $q_1 = 0.05$

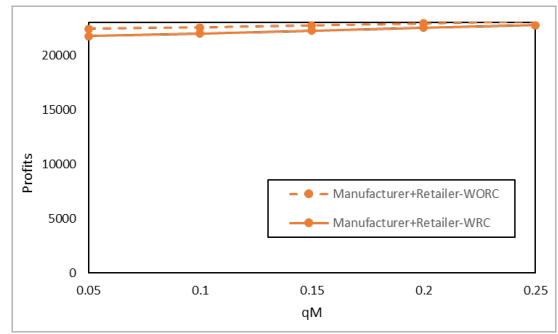
(b) $q_1 = 0.25$

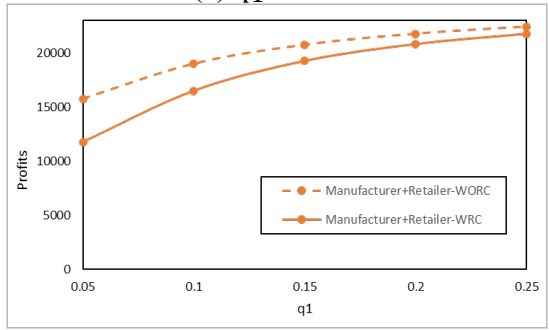
(c) $q_M = 0.05$

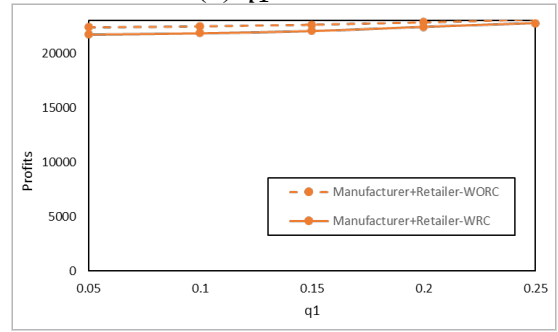
(d) $q_M = 0.25$

Figure 17. Impact of retail competition on profit of the cooperative channel.

From Figures 16 and 17, we observe that both the individual members' profits and the total profit of the cooperative channel under the WRC scenario are significantly lower than those under the WORC scenario. When one member in the cooperative alliance has a low quality score (e.g., $q_1 = 0.05$), the negative impact of retail competition on profits gradually diminishes as its partner's quality score increases. Conversely, when one member has a high quality score (e.g.,

$q_M = 0.25$), the alliance's profit is less sensitive to changes in their partner's quality score. This pattern may be attributed to the fact that the members with high quality score typically possess stronger market positions and customer bases, which helps maintain alliance profitability even under competitive pressure.

## 7. Conclusions

In this paper, we explore the optimal cooperative advertising strategies in the search engine advertising (SEA). We consider two primary forms of cooperative games: scenarios with and without retail competition. Specifically, we first introduce a dynamic model framework with a finite planning horizon for cooperative advertising strategies. Given the specific characteristics of the SEA market, we incorporate quality scores to extend the Vidale-Wolfe model. Then, we propose a feedback equilibrium solution of optimal advertising policies for the manufacturer and retailers and present a computational solution to derive concrete optimal subsidy rates for the manufacturer. Furthermore, we conduct numerical experiments to evaluate how the quality score affects optimal advertising strategies for the retailers and the manufacturer, the corresponding market share trajectories and the two optimal subsidy rates in both scenarios. Additionally, we investigate the impact of the initial market share of the competing retailer in Scenario II. Finally, we conduct a comparative analysis to study the impact of retail competition on the cooperative alliance's optimal strategies and channel performance. Our findings indicate that, in different competitive scenarios, our cooperative advertising strategies effectively help Manufacturer-Retailer alliance achieve optimal profits in dynamic SEA markets. We also state the significant role of quality scores in moderating retail competition effects on optimal cooperative advertising strategies. Specifically, retail competition drives increased advertising efforts and subsidy rates, with higher quality scores helping to stabilize cooperative advertising strategies and mitigate negative competitive impacts on SEA market performance.

**7.1 Theoretical contributions**

This study makes significant theoretical contributions in four aspects.

First, we build a dynamic model framework to study the cooperative advertising strategies in the SEA market in a finite planning horizon. Prior research primarily utilized static models to construct strategic frameworks in cooperative advertising (e.g., Ahmadi-Javid and Hoseinpour, 2012; He et al., 2013). Recent studies have increasingly adopted dynamic models to more accurately reflect temporal changes. However, many of these studies use an infinite planning horizon to simplify their analyses (e.g., Bai et al., 2018; Ezimadu, 2019), which fails to capture the period updates that are characteristic of the SEA market. Our model framework not only addresses this limitation but also suggests potential applicability to other digital advertising forms.

Second, we develop a structural framework for analyzing cooperative advertising games under two different competitive scenarios. Through this framework, we have derived optimal analytical solutions for cooperative advertising strategies, significantly enhancing the understanding of the complex competition environment in the SEA.

Third, our model uniquely incorporates dynamic quality scores when optimizing cooperative strategies in SEA and conducts a sensitivity analysis on it. This analysis demonstrates that advertisers with higher quality scores achieve greater market shares and profits, and can further improve their competitiveness and market share against high quality score competitors by increasing their own quality scores and subsidizing retailers more heavily. Additionally, we also conduct sensitivity analysis on gross margins, which reveals that higher gross margins make manufacturer increase both its own advertising effort and subsidy rate to retailer to enhance the effectiveness and competitiveness of its advertising.

Finally, our analysis of the impact of retail competition reveals the quality score based complementarity effect in Manufacturer-Retailer alliances, where improvements in one member's quality score can compensate for their partner's competitive disadvantage, thereby reducing their sensitivity to retail competition in cooperative advertising decisions in SEA.

### 7.2 Practical implications

Our research provides essential managerial insights for advertisers engaged in SEA, demonstrating how quality score and gross margin significantly influence the optimal cooperative advertising strategies. These factors are critical, as they impact market share, profits, and subsidy rates across various cooperative scenarios in the SEA market.

First, in scenario without a competing retailer, the quality score of each member within the cooperative supply chain plays a crucial role in determining the manufacturer's subsidy rate. Therefore, it is important for advertisers to meticulously assess both their own and their partners' quality scores when determining their subsidy strategies. For example, the manufacturer tends to increase its subsidy rate for cooperative retailer when its quality score is high, aimed at motivating both itself and the cooperative retailer to increase their advertising efforts, thus helping to stabilize or expand the market share of their alliance. However, a manufacturer might reduce its subsidy rate when the cooperative retailer's quality score is high, and allocates more resources to improve its own advertising quality score.

Second, in scenario with competing retailers, besides the quality scores, advertisers should also consider their gross margins and the initial market shares. Specifically, when facing competitors with high quality scores and gross margins, manufacturers should enhance their competitiveness and profitability in the fiercely competitive SEA market by boosting advertising cooperation, such as increasing subsidy rates or advertising efforts. Moreover, manufacturers should also consider the initial market shares of both parties in competitive scenario. If the Manufacturer-Retailer alliance has a higher initial market share, the manufacture should increase the subsidy rate for the cooperative retailer to maintain its market share and profit; while if the competitor has a higher initial market share, the manufacturer should increase its advertising efforts at the initial stage to strengthen its own competitiveness and that of the Manufacturer-Retailer alliance.

Last but not least, our comparative analysis suggests that to enhance competitive advantage, manufacturers should not only improve their own quality score but also support their partners' quality score improvements through increased subsidy rates. Retailers should strategically adjust their advertising efforts based on both the subsidy support and their quality score position in the alliance. By leveraging the quality score based complementarity effect, alliances can better maintain their market performance under retail competition in the SEA market and achieve sustainable profit growth.

### 7.3 Future research

In the future, we plan to incorporate budget constraints to better reflect the resource limitations for both manufacturers and retailers. This will enable a more complex, yet realistic simulation of their advertising decisions. Moreover, to establish more flexible cooperative advertising strategies model and broaden the applicability of our findings in digital marketing contexts, we will adapt our cooperative advertising models to other forms of online advertising by incorporating their unique advertising characteristics and sophisticated interaction mechanisms.

**Appendix A**

*A1. Proof of Proposition 1*

The Hamilton-Jacobi-Bellman (HJB) equation for optimal control problems given by Equations (7) and (8) are

*Retailer-1:*

$$V_{1t} + \max_{u_1 \geq 0}\{e^{-rt}[c_1 x - (1-\theta)u_1^2] + V_{1x}\left((\rho_1 q_1(t)u_1(t) + \rho_M q_M(t)v(t))\sqrt{1-x}\right)\} = 0, \quad (A1.1)$$

*Manufacturer:*

$$V_{Mt} + \max_{v \geq 0}\{e^{-rt}[c_M x - v^2 - \theta u_1^2] + V_{Mx}\left((\rho_1 q_1(t)u_1(t) + \rho_M q_M(t)v(t))\sqrt{1-x}\right)\} = 0, \quad (A1.2)$$

where $V_1(x,T) = 0$, $V_M(x,T) = 0$, $V_{1t} = \partial V_1/\partial t$, $V_{Mt} = \partial V_M/\partial t$, $V_{1x} = \partial V_1/\partial x$, $V_{Mx} = \partial V_M/\partial x$.

Taking the first derivative of (A1.1) with respect to $u_1$ yields

$$-2(1-\theta)u_1 + \rho_1 q_1(t)V_{1x}\sqrt{1-x} = 0,$$

and taking the first derivative of (A1.2) with respect to $v$ yields

$$-2v + \rho_M q_M(t)V_{Mx}\sqrt{1-x} = 0,$$

which lead to the optimal advertising efforts of Retailer-1 and the manufacturer as follows:

$$u_1^*(x,\theta) = \frac{\rho_1 q_1(t) e^{rt} V_{1x} \sqrt{1-x}}{2(1-\theta)}, \tag{A1.3}$$

$$v^*(x,\theta) = \frac{\rho_M q_M(t) e^{rt} V_{Mx} \sqrt{1-x}}{2}. \tag{A1.4}$$

Substituting (A1.3) - (A1.4) into (A1.1) yields

$$V_{1t} + c_1 e^{-rt} x + \frac{(\rho_1 q_1(t))^2 e^{rt} V_{1x}^2 (1-x)}{4(1-\theta)} + \frac{(\rho_M q_M(t))^2 e^{rt} V_{1x} V_{Mx} (1-x)}{2} = 0, \tag{A1.5}$$

$$V_{Mt} + c_M e^{-rt} x + \frac{(\rho_M q_M(t))^2 e^{rt} V_{Mx}^2 (1-x)}{4} - \frac{(\rho_1 q_1(t))^2 \theta e^{rt} V_{1x}^2 (1-x)}{4(1-\theta)^2}$$

$$+ \frac{(\rho_1 q_1(t))^2 e^{rt} V_{1x} V_{Mx} (1-x)}{2(1-\theta)} = 0. \tag{A1.6}$$

### A2. Proof of Proposition 2

Conjecture the following functional forms for the value functions $V_1(x,t)$ and $V_M(x,t)$:

$$V_1(x,t) = e^{-rt}(\alpha_1(t) + \beta_1(t) x),$$

$$V_M(x,t) = e^{-rt}(\alpha_M(t) + \beta_M(t) x),$$

where $\alpha_1(t), \beta_1(t), \alpha_M(t), \beta_M(t)$ are time-dependent functions to be determined. Given the terminal value conditions $V_1(x,T) = 0$ and $V_M(x,T) = 0$, we have $\alpha_1(T) = 0$, $\beta_1(T) = 0$, $\alpha_M(T) = 0$, and $\beta_M(T) = 0$.

Then we can easily obtain

$$V_{1t} = -re^{-rt}((\alpha_1(t) + \beta_1(t) x)) + e^{-rt}((\dot{\alpha}_1(t) + \dot{\beta}_1(t) x)), \tag{A2.1}$$

$$V_{1x} = e^{-rt} \beta_1(t), \tag{A2.2}$$

$$V_{Mt} = -re^{-rt}(\alpha_M(t) + \beta_M(t) x) + e^{-rt}((\dot{\alpha}_M(t) + \dot{\beta}_M(t) x)), \tag{A2.3}$$

$$V_{Mx} = e^{-rt} \beta_M(t). \tag{A2.4}$$

Substituting (A2.1), (A2.2) and (A2.4) into (A1.5) and by further calculating, we have

$$\dot{\alpha}_1 = r\alpha_1 - \frac{(\rho_1 q_1(t))^2 \beta_1^2}{4(1-\theta)} - \frac{(\rho_M q_M(t))^2 \beta_1 \beta_M}{2}, \alpha_1(T) = 0,$$

$$\dot{\beta}_1 = r\beta_1 + \frac{(\rho_1 q_1(t))^2 \beta_1^2}{4(1-\theta)} + \frac{(\rho_M q_M(t))^2 \beta_1 \beta_M}{2} - c_1, \beta_1(T) = 0.$$

Substituting (A2.2), (A2.3) and (A2.4) into (A1.6) and by further calculating, we have

$$\dot{\alpha}_M = r\alpha_M - \frac{(\rho_1 q_1(t))^2 \beta_M^2}{4} + \frac{(\rho_1 q_1(t))^2 \theta \beta_1^2}{4(1-\theta)^2} - \frac{(\rho_1 q_1(t))^2 \beta_1 \beta_M}{2(1-\theta)}, \alpha_M(T) = 0,$$

$$\dot{\beta}_M = r\beta_M + \frac{(\rho_M q_M(t))^2 \beta_M^2}{4} - \frac{(\rho_1 q_1(t))^2 \theta \beta_1^2}{4(1-\theta)^2} + \frac{(\rho_1 q_1(t))^2 \beta_1 \beta_M}{2(1-\theta)} - c_M, \beta_M(T) = 0.$$

Then, we can derive the optimal advertising efforts of Retailer-1 and the manufacturer as

$$u_1^* = \frac{\rho_1 q_1(t)\beta_1(t)\sqrt{1-x}}{2(1-\theta)}, \tag{A2.5}$$

$$v^* = \frac{\rho_M q_M(t)\beta_M(t)\sqrt{1-x}}{2}. \tag{A2.6}$$

Substituting (A2.5) and (A2.6) into Equation (1) yields

$$\dot{x} = \left((\rho_1 q_1(t))^2 \frac{\beta_1(t)}{2(1-\theta)} + (\rho_M q_M(t))^2 \frac{\beta_M(t)}{2}\right)(1-x),\ x(0) = x_0.$$

### A3. Proof of Proposition 4

The Hamilton-Jacobi-Bellman (HJB) equation for optimal control problems given by Equations (30) and (31) are

*Retailer-1:*

$$V_{1t} + \max_{u_1 \geq 0}\{e^{-rt}[c_1 x - (1-\theta)u_1^2] + V_{1x}\left((\rho_1 q_1(t)u_1(t) + \rho_M q_M(t)v(t))\sqrt{1-x(t)} - \rho_2 q_2(t)u_2(t)\sqrt{x(t)}\right)\} = 0, \tag{A3.1}$$

*Retailer-2:*

$$V_{2t} + \max_{u_2 \geq 0}\{e^{-rt}[c_2(1-x) - u_2^2] + V_{2x}\left((\rho_1 q_1(t)u_1(t) + \rho_M q_M(t)v(t))\sqrt{1-x(t)} - \rho_2 q_2(t)u_2(t)\sqrt{x(t)}\right)\} = 0, \tag{A3.2}$$

*Manufacturer:*

$$V_{Mt} + \max_{v \geq 0}\{e^{-rt}[c_M x - v^2 - \theta u_1^2] + V_{Mx}\left((\rho_1 q_1(t)u_1(t) + \rho_M q_M(t)v(t))\sqrt{1-x(t)} - \rho_2 q_2(t)u_2(t)\sqrt{x(t)}\right)\} = 0, \tag{A3.3}$$

where $V_1(x,T) = 0$, $V_2(x,T) = 0$, $V_M(x,T) = 0$, $V_{1t} = \partial V_1/\partial t$, $V_{2t} = \partial V_2/\partial t$, $V_{Mt} = \partial V_M/\partial t$, $V_{1x} = \partial V_1/\partial x$, $V_{2x} = \partial V_2/\partial x$, $V_{Mx} = \partial V_M/\partial x$.

Taking the first derivative of (A3.1) with respect to $u_1$ and the first derivative of (A3.2) with respect to $u_2$ yields

$$-2(1-\theta)u_1 + \rho_1 q_1(t)V_{1x}\sqrt{1-x} = 0,$$

$$-2u_2 - \rho_2 q_2(t)V_{2x}\sqrt{x(t)} = 0,$$

and taking the first derivative of (A3.3) with respect to $v$ yields

$$-2v + \rho_M q_M(t)V_{Mx}\sqrt{1-x} = 0,$$

which lead to the optimal advertising efforts of Retailer-1, Retailer-2 and the manufacturer as follows:

$$u_1^* = \frac{\rho_1 q_1(t)) V_{1x}\sqrt{1-x}}{2(1-\theta)}, \tag{A3.4}$$

$$u_2^* = -\frac{\rho_2 q_2(t) V_{2x}\sqrt{x}}{2}, \tag{A3.5}$$

$$v^* = \frac{\rho_M q_M(t) V_{Mx}\sqrt{1-x}}{2}. \tag{A3.6}$$

Conjecture the following functional forms for the value functions $V_1(x,t)$, $V_2(x,t)$ and $V_M(x,t)$:

$$V_1(x,t) = e^{-rt}(\alpha_1(t) + \beta_1(t)x), \tag{A3.7}$$

$$V_2(x,t) = e^{-rt}(\alpha_2(t) + \beta_2(t)x), \tag{A3.8}$$

$$V_M(x,t) = e^{-rt}(\alpha_M(t) + \beta_M(t)x), \tag{A3.9}$$

where $\alpha_1(t), \beta_1(t), \alpha_2(t), \beta_2(t), \alpha_M(t), \beta_M(t)$ are time-dependent functions to be determined. Given the terminal value conditions $V_1(x,T) = 0$, $V_2(x,T) = 0$ and $V_M(x,T) = 0$, we have $\alpha_1(T) = 0$, $\beta_1(T) = 0$, $\alpha_2(T) = 0$, $\beta_2(T) = 0$, $\alpha_M(T) = 0$, and $\beta_M(T) = 0$.

Then we can easily obtain

$$V_{1x} = e^{-rt}\beta_1(t) \tag{A3.10}$$

$$V_{2x} = e^{-rt}\beta_2(t) \tag{A3.11}$$

$$V_{Mx} = e^{-rt}\beta_M(t) \tag{A3.12}$$

Substituting (A3.10) into (A3.4), (A3.11) into (A3.5) and (A3.12) into (A3.6) yield

$$u_1^* = \frac{\rho_1 q_1(t)\beta_1(t)\sqrt{1-x}}{2(1-\theta)} \tag{A3.13}$$

$$u_2^* = -\frac{\rho_2 q_2(t)\beta_2(t)\sqrt{x}}{2} \tag{A3.14}$$

$$v^* = \frac{\rho_M q_M(t)\beta_M(t)\sqrt{1-x}}{2} \tag{A3.15}$$

Substituting (A3.7), (A3.8), (A3.9), (A3.13), (A3.14) and (A3.15) into (A3.1), (A3.2) and (A3.3) by further calculating, we have

$$\dot{\alpha}_1 = r\alpha_1 - \frac{(\rho_1 q_1(t))^2 \beta_1^2}{4(1-\theta)} - \frac{(\rho_M q_M(t))^2 \beta_1 \beta_M}{2}, \alpha_1(T) = 0,$$

$$\dot{\beta}_1 = r\beta_1 + \frac{(\rho_1 q_1(t))^2 \beta_1^2}{4(1-\theta)} - \frac{(\rho_2 q_2(t))^2 \beta_1 \beta_2}{2} + \frac{(\rho_M q_M(t))^2 \beta_1 \beta_M}{2} - c_1, \beta_1(T) = 0,$$

$$\dot{\alpha}_2 = r\alpha_2 - \frac{(\rho_1 q_1(t))^2 \beta_1 \beta_2}{2(1-\theta)} - \frac{(\rho_M q_M(t))^2 \beta_1 \beta_M}{2} - c_2, \alpha_2(T) = 0,$$

$$\dot{\beta}_2 = r\beta_2 + \frac{(\rho_1 q_1(t))^2 \beta_1 \beta_2}{2(1-\theta)} + \frac{(\rho_M q_M(t))^2 \beta_2 \beta_M}{2} - \frac{(\rho_2 q_2(t))^2 \beta_2^2}{4} + c_2, \beta_2(T) = 0,$$

$$\dot{\alpha}_M = r\alpha_M - \frac{(\rho_M q_M(t))^2 \beta_M^2}{4} + \frac{(\rho_1 q_1(t))^2 \theta \beta_1^2}{4(1-\theta)^2} - \frac{(\rho_1 q_1(t))^2 \beta_1 \beta_M}{2(1-\theta)}, \alpha_M(T) = 0,$$

$$\dot{\beta}_M = r\beta_M + \frac{(\rho_M q_M(t))^2 \beta_M^2}{4} - \frac{(\rho_1 q_1(t))^2 \theta \beta_1^2}{4(1-\theta)^2} + \frac{(\rho_1 q_1(t))^2 \beta_1 \beta_M}{2(1-\theta)} - \frac{(\rho_2 q_2(t))^2 \beta_2 \beta_M}{2} - c_M, \beta_M(T) = 0.$$

Substituting (A3.13), (A3.14) and (A3.15) into Equation (2) yields

$$\dot{x} = \left((\rho_1 q_1(t))^2 \frac{\beta_1(t)}{2(1-\theta)} + (\rho_M q_M(t))^2 \frac{\beta_M(t)}{2}\right)(1-x) + \frac{(\rho_2 q_2(t))^2 \beta_2(t) x}{2}, x(0) = x_0.$$